\documentclass[aps,prl,reprint,superscriptaddress,nofootinbib]{revtex4-1}

\usepackage{amsmath,amssymb,amsfonts,dsfont,graphicx}

\newcommand{\A}{{\cal A}}
\def\r{{\bf r}}

\usepackage{tikz}

\newcommand*\circled[1]{\tikz[baseline=(char.base)]{
    \node[shape=circle, draw, inner sep=1pt,
        minimum height=12pt] (char) {#1};}}

\font\upright=cmu10 scaled\magstep1
\def\stroke{\vrule height8pt width0.4pt depth-0.1pt}
\def\Cmath{\vcenter{\hbox{\upright\rlap{\rlap{C}\kern
               3.8pt\stroke}\phantom{C}}}}
\def\Rmath{ {\bf R}}
\def\Hmath{\vcenter{\hbox{\upright\rlap{\rlap{H}\kern
                   3.8pt\stroke}\phantom{H}}}}

\def\R{\ifmmode\Rmath\else$\Rmath$\fi}

\def\barray{\begin{eqnarray}}
\def\earray{\end{eqnarray}}
\def\beq{\begin{equation}}
\def\eeq{\end{equation}}

\def\x{{\bf x}}
\begin{document}

\title{Field Tensor Network States}

\author{Anne E. B. Nielsen}
\altaffiliation{On leave from Department of Physics and Astronomy, Aarhus University, DK-8000 Aarhus C, Denmark.}
\affiliation{Max-Planck-Institut f\"ur Physik komplexer Systeme, N{\"o}thnitzer Str.\ 38, D-01187 Dresden, Germany}

\author{Benedikt Herwerth}
\affiliation{Max-Planck-Institut f\"ur Quantenoptik, Hans-Kopfermann-Str.\ 1, D-85748 Garching, Germany}

\author{J. Ignacio Cirac}
\affiliation{Max-Planck-Institut f\"ur Quantenoptik, Hans-Kopfermann-Str.\ 1, D-85748 Garching, Germany}
\affiliation{Munich Center for Quantum Science and Technology (MCQST), Schellingstr.\ 4, D-80799 M{\"u}nchen}

\author{Germ\'an Sierra}
\affiliation{Instituto de F\'{\i}sica Te\'orica  UAM/CSIC, Universidad Aut\'onoma de Madrid, Cantoblanco, Madrid, Spain}

\begin{abstract}
We define a class of tensor network states for spin systems where the individual tensors are functionals of fields. The construction is based on the path integral representation of correlators of operators in quantum field theory. These tensor network states are infinite dimensional versions of matrix product states and projected entangled pair states. We find the field-tensor that generates the Haldane-Shastry wave function and extend it to two dimensions. We give evidence that the latter underlies the topological chiral state described by the Kalmeyer-Laughlin wave function.
\end{abstract}

\maketitle

Tensor networks (TN) are becoming a key tool to describe many-body quantum systems \cite{RO19}. On the one hand, they can efficiently approximate quantum states of local Hamiltonians in thermal equilibrium, which has led to powerful numerical algorithms with applications in condensed matter and, to some extent, in high-energy physics \cite{Breview}. On the other hand, they provide us with paradigmatic examples of strongly correlated states and thus allow us to investigate intriguing many-body quantum phenomena. For instance, they offer us a guide to classify symmetry protected topological phases \cite{C11,S11}, or to understand a large variety of topologically ordered behavior. In fact, states (or models) like the AKLT \cite{AKLT}, string-net states \cite{L04}, or resonating valence-bond states have a very simple description in terms of TN. By simple we mean with a small bond dimension, $D$, which limits the number of coefficients describing the tensors generating the many-body states. The description of such states in terms of TN automatically opens up the possibility of using powerful tools in order to describe their physical properties by just inspecting a simple tensor. In 1D, one can easily describe symmetries and string order parameters \cite{P08}, or even gauge symmetries \cite{K17}. In 2D, apart from obtaining the physical symmetries, one can directly identify the topological properties or type of anyon excitations of the parent Hamiltonian \cite{W17}.

There exist, however, some classes of states for which no exact expressions in terms of tensor network states of finite bond dimensions exist. Two prominent examples are critical states \cite{M10}, and chiral topological states of gapped Hamiltonians in one and two dimensional spin lattices, respectively \cite{K87,W89}. The reason behind the lack of description as TN for the first stems from the fact that critical states violate the area law  \cite{area1,area3}. Specifically, the entanglement entropy of a connected region containing $L$ spins scales as $\propto \ln(L)$ \cite{V03,CC04}, whereas for a matrix product state (MPS), the one-dimensional version of tensor network states, it is bounded by $2\ln(D)$; therefore, in the thermodynamic limit for any finite $D$, there always exists some $L$ for which an MPS cannot cope with the amount of entanglement and thus it is impossible that it describes a critical state. The reason for the second class is more subtle and yet not fully understood; however, there are good reasons to believe that there exist obstructions due to the non-existence of local Wannier states \cite{fidkowski} (see, however \cite{B11}). In fact, for Gaussian fermionic states, it is not possible to describe gapped chiral topological insulators \cite{D15}\footnote{Even though those do not posses topological order, but rather belong to a symmetry protected topological phase, they constitute a strong evidence for the impossibility of describing chiral topological phases corresponding to gapped local Hamiltonians.}. We emphasize that here we mean an exact description; in fact, both classes of states may well be approximated efficiently with an error that decreases as $D$ increases
\cite{Z12,E13a,E13b}.

The arguments above do not prevent the existence of exact descriptions of critical or chiral topological states with TN of infinite bond dimensions. In \cite{C10}, it was noted that the conformal field theory (CFT) formulation \cite{M91} of the Haldane-Shastry state has similarities with MPS, and in \cite{Z12,E13a,E13b}, the CFT formulation was used to obtain MPS with a discrete, infinite bond dimension describing chiral topological states in 2D. From the tensor network perspective it is, however, desirable to use projected entangled pair states (PEPS) to deal with 2D systems. Furthermore, although the approach of \cite{Z12,E13a,E13b} can, in principle, be used to describe critical states in 1D with open boundary conditions, it is more appropriate to use periodic boundary conditions for translationally invariant systems.

In this letter we define Field Tensor Networks (FTN) for spin lattices in any dimension, where the bonds in the tensors are functions, the corresponding contractions are accomplished by a path integration, and the tensors themselves are functionals. The virtual space is hence continuous. We show how this approach can be used to describe translationally invariant critical systems, as well as the analogs of PEPS for two dimensional systems. Our construction is reminiscent of recent proposals for constructing tensor networks for quantum fields, where path integration is also employed \cite{J15,T18}. In our case, however, we deal with discrete spin lattices and the construction is quite different. We give a procedure to compute the FTN for states whose coefficients in the spin basis can be written as vacuum correlators of a quantum field theory with a local action. In particular, we give an explicit construction for free boson CFTs and vertex operators. We also take advantage of the fact that the Haldane-Shastry state \cite{H88,S88}, a prominent critical state, can be expressed in that form \cite{C10,N11} to compute a FTN generating that state. The description allows both periodic and open boundary conditions. We also propose a FTN in 2D and give strong evidence that it represents a Kalmeyer-Laughlin state \cite{K87}, a prototypical representative of chiral topological order.

{\em FTN in 1D:} We consider a spin chain of $N$ spins of dimension $d$, and a translationally invariant state
\begin{equation}
 |\Psi\rangle = \sum_{s_1,\ldots,s_N=1}^d c_{s_1,\ldots,s_N} |s_1,\ldots,s_N\rangle.
\end{equation}
Let us start recalling MPS, where
\begin{equation}
 c_{s_1,\ldots,s_N} = \sum_{n_1,\ldots,n_N=1}^D A^{s_1}_{n_1,n_2} \ldots A^{s_N}_{n_N,n_1},
\end{equation}
$D$ is the bond dimension, and for each value of $s=1,\ldots, d$, $A^s_{n,m}$ is a $D\times D$ matrix. In an analogous way we define  (translationally invariant) Field Tensor Network States (FTNS) as
\begin{equation}\label{FTNS}
c_{s_1,\ldots,s_N} = \int {\cal D}[\alpha_1]\ldots {\cal D}[\alpha_N] \A^{s_1}_{\alpha_1,\alpha_2}  \ldots \A^{s_N}_{\alpha_N,\alpha_1} \, .
\end{equation}
Here, $\alpha_n:\mathds{R}\to \mathds{R}$ belong to the set of square-integrable functions and also include a constant function, and for each value of $s=1,\ldots,d$, $\A^s_{\alpha,\beta}$ are functionals of $\alpha,\beta$. Note that $\Psi$ has the same structure as a MPS, where the indices of the matrices are replaced by the functions $\alpha_n$, and the sum over repeated indices is replaced by a path integral.

We also define the functionals
\begin{equation}\label{allA}
\A^{s_1,\ldots,s_n}_{\alpha,\alpha'} = \int {\cal D}[\alpha_2]\ldots {\cal D}[\alpha_n] \A^{s_1}_{\alpha,\alpha_2}  \ldots \A^{s_n}_{\alpha_n,\alpha'}
\end{equation}
fulfilling
\begin{subequations}\label{sewclos}
\begin{eqnarray}
\A^{s_1,\ldots,s_{n+1}}_{\alpha,\alpha'} &=& \int D[\beta] \A^{s_1,\ldots,s_{n}}_{\alpha,\beta} \A^{s_{n+1}}_{\beta,\alpha'},\\
c_{s_1,\ldots,s_N} &=& \int D[\alpha] \A^{s_1,\ldots,s_N}_{\alpha,\alpha},
\end{eqnarray}
\end{subequations}
which we will call ``sewing'' and ``closing'' conditions.

{\em Example:} Particularly interesting examples of FTNS are those for which the coefficient can be written in terms of correlators of a simple CFT in 1+1 dimensions and a local action. Specifically, we shall here study a family of critical states in 1D with
\begin{multline}\label{c1}
c_{s_1,\ldots,s_N}\propto \\ \delta_{\sum_n s_n,0} \prod_n \chi_{s_n} \prod_{n >  m} \left\{ \sin[(x_n-x_m)/N] \right\}^{2q^2  s_n s_m}.
\end{multline}
Here, $x_n=\pi (n- \frac{1}{2}) $, $q$ is a real number, $\chi_{s_n}$ is a phase factor that may depend on $s_n$, and $s_n=\pm1$. The member of the family with $q=1/2$ and $\chi_{s_n}=e^{in\pi (s_n-1)/2}$ is the ground state of the Haldane-Shastry model, which has been extensively studied in the literature as a paradigm of criticality. The wave function (\ref{c1}) has also been  used as an ansatz for the ground state of the  Hamiltonian of the XXZ  spin 1/2 chain in the critical regime with anisotropy parameter $\Delta = - \cos( 4 \pi q^2)$ \cite{C10}.

The states \eqref{c1} violate the area law, so that they cannot be written as MPS with finite bond dimension. Nevertheless, we will show here how they can be expressed as FTNS.

\begin{figure}
\includegraphics[width=\columnwidth,trim=44mm 122mm 47mm 110mm]{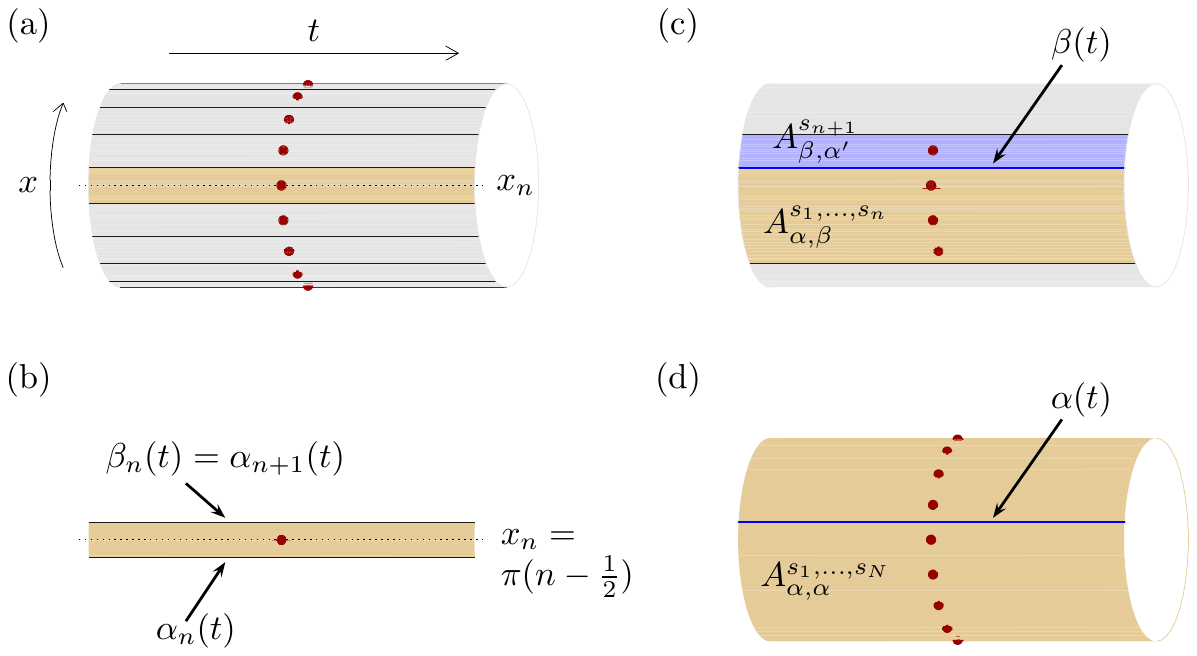}
\caption{(a) The CFT is defined on a cylinder of radial (axial) coordinate $x$ ($t$). The spins (represented as red points) are punctures at positions $(x_n,0)$. (b) The path integral defining the state is extended over the cylinder. The functional $A^{s_n}_{\alpha_n,\alpha_{n+1}}$ corresponds to the integration over a stripe, where the boundary conditions for $\varphi$ are given by $\alpha_n$ and $\alpha_{n+1}$. Panels (c) and (d) illustrate the sewing and closing conditions, respectively.}\label{Fig1}
\end{figure}

It is not difficult to show that \cite{CFT}
\begin{equation}\label{QFT}
c_{s_1,\ldots,s_N}\propto\langle \chi_{s_1} :e^{i  q   s_1 \varphi(\r_1)}:  \ldots \chi_{s_N} :e^{i  q   s_N \varphi(\r_N)}:  \rangle_0,
\end{equation}
where $\varphi$ is a real scalar field defined on a cylinder of circumference $\pi N$, $::$ denotes normal ordering, the $\r_n=(x_n,0)$ are points in cylindrical coordinates (see Fig.\ \ref{Fig1}), and the expectation value is taken in the vacuum. In this case, using the path integral representation we have
\begin{equation}
\label{cQFT}
c_{s_1,\ldots,s_N}\propto\int D[\varphi] e^{-S} e^{i q \sum_{n=1}^N s_n \varphi(\r_n)} \prod_n \chi_{s_n},
\end{equation}
where
\begin{equation}\label{S}
S = \frac{1}{8 \pi}  \int_0^{\pi N } dx \int_{-\infty}^\infty dt    \{[\partial_x \varphi(x,t) ]^2 +   [\partial_t \varphi(x,t) ]^2 \} \,
\end{equation}
is the Euclidean action of the boson field. Notice that (\ref{S}) vanishes if $\varphi$ is a constant $\varphi_0$ which upon integration generates the constraint $\sum_n s_n=0$ appearing in (\ref{c1}).

In order to find the FTNS representation of (\ref{QFT}), we rewrite
\begin{equation}
\int D[\varphi] =  \int D[\alpha_1] \ldots D[\alpha_N]
\int' D[\varphi_{1}] \ldots \int' D[\varphi_{N}].
\end{equation}
Here, $\alpha_n(t)$ is a function of $t$ only, and $\varphi_{n}(x,t)$ is defined in the interval $(x,t)\in [x_n -\delta,x_n + \delta]\times \mathds{R}$ with  $\delta=\pi/2$, and the $'$ indicates that it fulfills the boundary condition (see Fig.\ \ref{Fig1})
\begin{subequations}
\begin{eqnarray}
\varphi_{n}(x_n-\delta,t)&=&\alpha_{n}(t),\\
\varphi_{n}(x_n+\delta,t)&=&\alpha_{n+1}(t).
\end{eqnarray}
\end{subequations}
Thus, we simply identify
\begin{equation}
\A^{s_1, \dots, s_L}_{\alpha,\alpha'}=  \int' D[\varphi] e^{-S} e^{i  q  \sum_{j=1}^L   s_j   \varphi(x_j,0)} \prod_{n=1}^L \chi_{s_n},
\label{Amp}
\end{equation}
where the $'$ indicates that $\varphi$ has $\alpha = \alpha_1$ and $\alpha' = \alpha_{L+1}$ as boundary conditions, and $S$ is defined as in (\ref{S}) but with the integral in $x$ restricted to the interval $[x_1-\delta,x_L+\delta]$. Note that (\ref{FTNS}) trivially follows. The sewing and closing conditions (\ref{sewclos}) are represented for this case in Fig.\ \ref{Fig1}(c,d).

In the supplemental material \cite{SM}, using Green's function techniques, we explicitly compute
\begin{equation}\label{Asa}
\A^{s_1, \dots, s_L}_{\alpha,\alpha'} \propto   e^{ i \varphi_0 q \sum_j s_j}
e^{ S_L^{(0)} +   S_L^{(1) } +  S_L^{(2)}   }  \prod_{n=1}^L \chi_{s_n},
\end{equation}
where
\begin{eqnarray}
 \label{SM1}
S_L^{(0)}  & = &  2 q^2 \sum_{L \geq n > m \geq 1} s_n s_m  \ln \frac{ \sin[  (x_n - x_m)/(2 L)]}{  \sin[ (x_n +  x_m)/(2 L)]}  \ ,
\\
\label{SM2}
S_L^{(1)}  & = &   \frac{1}{64 \pi^2}  \int_\mathds{R} dt  dt'  \, \vec \alpha(t) U_L(t-t') \vec\alpha(t')^T  \, ,
  \\
  \label{SM3}
  S_L^{(2)} &  = &  \frac{q }{4 \pi}
  \int_\mathds{R} dt \, \sum_{n=1}^L  s_n    \vec \alpha(t)  (  \vec v_{L,n}(t)^T -  \vec v_{L,n}(t)^\dagger ) \, .
\end{eqnarray}
and $\varphi_0$ is the zero mode of the boson field that has been subtracted from the functions $\alpha, \alpha'$ to guarantee their normalizability. Here $\vec\alpha=(\alpha,\alpha')$, $U$ is a $2\times 2$ matrix with elements
\begin{eqnarray}
\label{u1}
 U_{L, 11 \;  {\rm or } \; 22 }(t)&=&     \frac{2}{L^2}  \left[
 \frac{1}{\sinh^2 \left( \frac{t}{ 2L} \right)}  - \left( \frac{ 2L}{t} \right)^2 \right]  - 8  P' \left(  \frac{1}{t}  \right)
 \nonumber
   \\
 U_{L,12 \; {\rm or} \; 21}(t)&=& \frac{2}{ L^2 \cosh^2 \left( \frac{ t}{2 L} \right) }
\end{eqnarray}
and
\begin{equation}
\vec v_{L,n}(t)= \frac{1}{L}   \left( \coth \left(\frac{ t- i x_n}{ 2 L}\right) , -  \tanh  \left(\frac{ t- i x_n}{ 2 L}\right) \right),
\end{equation}
where  $P' \left(  \frac{1}{t}  \right)= - \frac{1}{2}  \left( \frac{1}{(t + i 0)^2} + \frac{1}{(t - i 0)^2} \right)$ is the derivative of the principal value distribution $P \left(  \frac{1}{t}  \right)$.

{\em Chiral version:} We aim at also being able to describe chiral states, and as a test case, we next consider a chiral formulation of the critical states \eqref{c1}. In this formulation the states are defined in terms of a chiral free boson field $\varphi(z)$, which depends on $z$, but not on its conjugate $\bar{z}$. The states are again given by (\ref{QFT}), except that the vertex operators now take the form $:e^{i  q s_n \varphi(z_n)}:$, where $q\in\mathds{R}$ and $z_n=t+ix_n$ (the  wave function obtained with these chiral vertex operators coincides with \eqref{c1} except that
$\sqrt{2} q$ is replaced by $q$). This correlator can be written as in (\ref{cQFT}) with a chiral action \cite{F87} employed to study the edge excitations in the Quantum Hall effect \cite{W91}. However, the slicing of the path integral into the intervals $(x,t) \in [x_n -\delta,x_n + \delta]\times \mathds{R}$ introduces boundaries that mix the left and right moving modes of the bosonic field, which in turn complicates the approach.

We notice, however, that in \eqref{SM3}  there are two parts related by complex conjugation. Moreover, \eqref{SM1} comes from a Green function with four terms where only one of them  is analytic in the location of the vertex operators. It is therefore natural to expect that one obtains the chiral state by selecting only one of those parts. We shall use this property to define the new tensors
\begin{equation}\label{hatA}
\hat{\A}^{s_1, \dots, s_L}_{\alpha,\alpha'}=  e^{ i \varphi_0  q \sum_j s_j}
L^{ -   \frac{1}{2} L q^2  }  e^{ \hat{S}_L^{(0)} +  S_L^{(1) } +   \hat{S}_L^{(2)}   } \prod_{n=1}^L \chi_{s_n}
\end{equation}
where
\begin{eqnarray}
\hat{S}_L^{(0)}  & = &    q^2 \sum_{L \geq n > m \geq 1} s_n s_m  \ln \left[ 2  \sin \left( \frac{ x_n - x_m}{2 L}\right)\right],  \label{SM4}\\
\hat{S}_L^{(2)} &  = &  \frac{q }{4 \pi}
\int_\mathds{R} dt \, \sum_{n=1}^L  s_n    \vec \alpha(t)   \vec v_{L,n}(t)^T  \, . \label{SM5}
\end{eqnarray}
The factor $L^{ - \frac{1}{2}  L q^2  }$ guarantees that (\ref{hatA}) satisfies the sewing condition (\ref{sewclos}a).

The wave function that one obtains using \eqref{hatA} coincides with \eqref{c1} (although with $q$ replaced by $\sqrt{2}q$, so that the Haldane-Shastry state now corresponds to $q = 1/\sqrt{2}$). This is not evident when comparing Eqs.\ (\ref{hatA}-\ref{SM5}) with (\ref{Asa}-\ref{SM3}) as they look very different, and it is not obvious that they are related by a gauge transformation \cite{RO19}. However, one can prove this statement by showing that \eqref{hatA} fulfills the sewing condition (\ref{sewclos}a) and that by closing (\ref{sewclos}b) one indeed obtains the Haldane-Shastry wavefunction. We show that in the supplementary material \cite{SM}, where we use the translational invariance of the action restricted to the strip along the $t$-coordinate, which allows us to diagonalize in $k$-space. Thus, the procedure leading to \eqref{hatA} provides a field theory version of the chiral vertex operator in CFT.

\begin{figure}
\includegraphics[width=\columnwidth,trim=44mm 140mm 47mm 128mm]{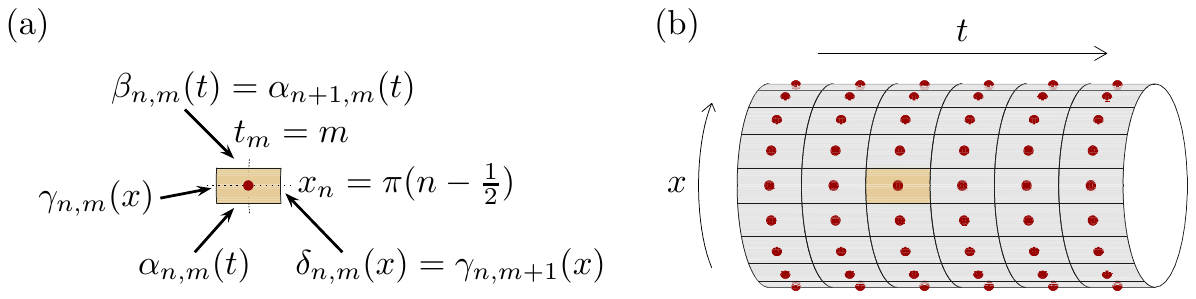}
\caption{FTNS in two dimensions: (a) functions on which $A^s_{\alpha_{n,m},\beta_{n,m},\gamma_{n,m},\delta_{n,m}}$ depends upon; (b) arrangement on a cylinder.}
\label{Fig2}
\end{figure}

{\em FTN in 2D:} The constructions presented above can be straightforwardly extended to represent states in two dimensions, corresponding to, e.g., a square spin lattice. We construct it on a cylinder, although one can similarly use a torus. The strip  $[x_n- \delta, x_n + \delta] \times \mathds{R}$ considered above is  replaced by the rectangle   $[x_n- \delta, x_n + \delta] \times [t_m - \delta', t_m + \delta']$. We define the functional $A^s_{\alpha_{n,m},\beta_{n,m},\gamma_{n,m},\delta_{n,m}}$, which depends on the functions $\alpha_{n,m},\beta_{n,m}:[t_m - \delta' ,t_m + \delta'] \to \mathds{R}$ and $\delta_{n,m},\gamma_{n,m}:[x_n-\delta,x_n+\delta] \to \mathds{R}$ (see Fig.\ \ref{Fig2}a). By arranging the functionals along the cylinder (Fig.\ \ref{Fig2}b), identifying functions like in Fig.\ \ref{Fig2}a, and integrating over them, one can construct states very much in the same way as one builds PEPS in two dimensions.

{\em Example:} Again, an illustrative example is provided by states that are expressed in terms of a simple CFT with a local action. For instance, we can define a spin state as (see Fig.\ 2)
\begin{equation}\label{as2d}
A^s_{\alpha_{n,m},\beta_{n,m},\gamma_{n,m},\delta_{n,m}} = \int' D[\varphi] e^{-S[\varphi]}  e^{i q s \varphi(x_n,t_m)}.
\end{equation}
Here, $\varphi:[x_n-\delta,x_n+\delta]\times [t_m- \delta' ,t_m+ \delta' ]\to \mathds{R}$, and with the boundary conditions
\begin{subequations}
\begin{eqnarray}
\varphi(x_n- \delta,t)&=&\alpha_{n,m}(t), \\
\varphi(x_n + \delta,t)&=&\beta_{n,m}(t),\\
\varphi(x, t_m - \delta' )&=&\gamma_{n,m}(x),\\
\varphi(x,  t_m +  \delta' )&=&\delta_{n,m}(x).
\end{eqnarray}
\nonumber
\end{subequations}
We then carry out the path integral in \eqref{as2d}. The main technical tool to do this is to apply a conformal map that transforms the rectangle $[x_n- \delta, x_n + \delta] \times [t_m - \delta', t_m + \delta']$ into the complex upper-half plane in terms of Jacobi elliptic functions. In the case $(n,m)=(1,0)$ one finds (see the supplemental material \cite{SM})

\begin{equation}\label{pep1}
A^{s}_{\alpha,\beta, \gamma, \delta}= e^{ S^{(1) } +   S^{(2)}   }
\end{equation}
where
\begin{eqnarray}
\label{pep2}
S^{(1)}  & = &   \frac{1}{64 \pi^2}  \int_{\cal I}  d\xi   d \xi'   \, \vec \alpha(\xi) U (\xi, \xi') \vec\alpha(\xi')^T  \, ,
\\
\label{pep3}
S^{(2)} &  = &  \frac{q  s}{4 \pi}
\int_{\cal I} d\xi  \,   \vec \alpha(\xi )  (  \vec v(\xi)^T -  \vec v(\xi)^\dagger ) \, ,
\end{eqnarray}
with $\vec \alpha = (\alpha, \beta, \gamma, \delta)$,  $\xi_1 = \xi_2 = t$ and $\xi_3 = \xi_4 = x$. The integration domain ${\cal I}$ is adapted to the type of variables  involved. $U$ is a $4 \times 4$ matrix  some of whose  elements are (the complete matrix is given in the supplemental material \cite{SM})
\begin{eqnarray}
& U_{11}(t, t') =  8   \left( \frac{  cn(t) dn(t) cn(t') dn(t') }{ (sn(t) -  sn(t'))^2 }- \frac{1}{(t-t')^2}  - P' \frac{1}{t-t'}    \right),   &  \nonumber   \\
& U_{33}(t, t') =  8  \left( \frac{k'^2   \widetilde{cn}(t)  \widetilde{sn}(t)  \widetilde{cn}(t') \tilde{sn}(t') }{ ( \widetilde{dn}(t) -  \widetilde{dn}(t'))^2 }- \frac{1}{(t-t')^2}  - P'  \frac{1}{t-t'}     \right),& \nonumber  \\
& U_{12}(t,t') = U_{21}(t,t')  =    8 k \frac{ cn(t) dn(t) cn(t') dn(t')}{ ( 1- k^2 sn(t) sn(t'))^2},   & \nonumber    \\
& U_{34}(t,t') = U_{43}(t,t')  =    8 k'^4  \,
 \frac{ \widetilde{cn} (t)  \widetilde{sn}(t)   \widetilde{cn}(t')  \widetilde{ sn}(t')}{ ( \widetilde{dn}(t) + \widetilde{dn}(t')   )^2},   & \nonumber\\
& U_{22}(t,t')=U_{11}(t,t'), \quad U_{44}(t,t')=U_{33}(t,t'). & \label{pep2d}
\end{eqnarray}
Here, $sn(t)$, $cn(t)$, and $dn(t)$ are Jacobi elliptic functions of modulus $k$ and $\widetilde{sn}(t)$, $\widetilde{cn}(t)$, and $\widetilde{dn}(t)$ of modulus $k'= \sqrt{1 - k^2}$, where $k$ is determined from the aspect ratio $\delta/\delta'$ of the considered rectangle \cite{SM}. In the limit $k \rightarrow 1$, the rectangle degenerates into a strip and we recover the matrix elements \eqref{u1}.

As in the previous example we can truncate this functional to a chiral one
\begin{equation}\label{pep4}
\hat{A}^{s}_{\alpha,\beta, \gamma, \delta}= \chi_s e^{  S^{(1) } +   \hat{S}^{(2)}   }
\end{equation}
where the phase factor $\chi_s$ can be chosen at will and
\begin{eqnarray}
\hat{S}^{(2)} &  = &  \frac{ q  s}{4 \pi}
\int_{\cal I} d\xi  \,   \vec \alpha(\xi )   \vec v(\xi)^T  \, .
\end{eqnarray}
We here consider the system on a cylinder, and to remove the virtual degrees of freedom on the boundaries, we take the rectangular regions for the boundary spins to go all the way to infinity. These boundary tensors can be obtained following the same approach as for the tensors in the bulk. We conjecture that sewing these amplitudes we get the wave function
\begin{equation}\label{kl}
c_{s_1, \dots, s_N}\propto \delta_{\sum_n s_n,0} \prod_n \chi_{s_n} \prod_{n >m} (z_n - z_m)^{ q^2 s_n s_m}
\end{equation}
with $z_n = t_n + i x_n$. When $q=1/\sqrt{2}$, Eq.\ (\ref{kl}) is a 2D topological state in the same universality class as the bosonic Laughlin state at filling fraction $1/2$, and the Kalmeyer-Laughlin wave function is obtained for $N \to \infty$ \cite{N12}.

{\em Conclusions:} We introduced a new class of TN constructed using functionals of fields that are contracted by means of the path integral of the functions defined on the links of the network.  These tensors satisfy sewing and closing conditions that are similar to those  employed in the construction of the scattering amplitudes in string theory \cite{string1,string2}.

We illustrate our approach using a massless boson in 2D that allows us to derive the Haldane-Shastry wave function that describes a critical state in the universality class given by the WZW model $SU(2)_1$. We also conjecture the field-tensor that generates the Kalmeyer-Laughlin state, which suggests that the chiral PEPS underlying topological chiral states in 2D require infinite bond dimension. The latter suggestion could be further studied by truncating the field variables to a finite number of modes in which case the field-tensor provides a PEPS with finite bond dimension. We have here focused on lattice states, but utilizing the techniques in \cite{T13} to approach the continuum limit of the states, one could similarly describe continuum states. The definition of field tensor network states applies equally well to other types of lattices than those considered here. Our approach also allows a way to study topological chiral states based on the symmetry properties of the field-tensors.

\vspace{0.5cm}

{\em Acknowledgements.} We would like to thank A. Gasull, E.  L\'opez, and A. Tilloy for conversations. J.I.C. is supported by the Deutsche Forschungsgemeinschaft (DFG, German Research Foundation) under Germany's Excellence Strategy EXC-2111 390814868, and by the European Union through the ERC grant QUENOCOBA, ERC-2016-ADG (Grant no. 742102). G.S. is supported by the Ministerio de Ciencia, Innovaci\'on y Universidades (grant PGC2018-095862-B-C21), the Comunidad de Madrid (grant QUITEMAD+ S2013/ICE-2801),   the ``Centro de Excelencia Severo Ochoa'' Programme (grant SEV-2016-0597) and the CSIC Research Platform on Quantum Technologies PTI-001.

\widetext

\pagebreak

\setcounter{equation}{0}
\setcounter{figure}{0}
\setcounter{page}{1}

\renewcommand{\theequation}{S\arabic{equation}}
\renewcommand{\thefigure}{S\arabic{figure}}

\begin{center}
\textbf{\large Supplemental Material}\\
\vspace{3mm}
Anne E. B. Nielsen$^{1,*}$, Benedikt Herwerth$^2$, J. Ignacio Cirac$^{2,3}$, and Germ\'an Sierra$^4$\\[1mm]
{\it\small
$^1$Max-Planck-Institut f\"ur Physik komplexer Systeme, N{\"o}thnitzer Str.\ 38, D-01187 Dresden, Germany\\
$^2$Max-Planck-Institut f\"ur Quantenoptik, Hans-Kopfermann-Str.\ 1, D-85748 Garching, Germany\\
$^3$Munich Center for Quantum Science and Technology (MCQST), Schellingstr.\ 4, D-80799 M{\"u}nchen\\
$^4$Instituto de F\'{\i}sica Te\'orica  UAM/CSIC, Universidad Aut\'onoma de Madrid, Cantoblanco, Madrid, Spain}
\end{center}

\section{General formalism}

Let $\phi(x_1, x_2)$ be a  real massless scalar field  in a simply connected region $M$ of
the two dimensional spacetime $\Rmath^2$.
The Euclidean free  action  of this field  is given  by
\beq
S_{M}^{(0)}(\phi)    =  \frac{1}{8 \pi}  \int_{M}  d^2 x  \;   \partial_{x_i}  \phi  \;  \partial_{x_i}   \phi  \, .
\label{a1}
\eeq
Let $f(\x)$ be the values that $\phi(\x)$ takes at the boundary $\partial M$ of the region $M$,
and   $ \{ q_j \}_{j=1}^N$ a set of $N$ charges  located at the  positions $\x_j \in M$
that corresponds to the   charge density
\beq
\rho=  4 \pi i  \sum_{j=1}^N  q_j   \delta^2(\x- \x_j) \,  \,.
\label{a12}
\eeq
We shall associate  to these  data the path integral
\beq
Z_{M}( f, \rho) =
 \int_{ \phi|_{ \partial {M}}  =  f }  [ d \phi]    \, e^{ - S_{M}(\phi)      },
\label{a2}
\eeq
with
\beq
S_{M}(\phi)   = S_{M}^{(0)}(\phi)
 -  \frac{1}{4 \pi}  \int_{M}  d^2 x  \; \phi  \,  \rho   \, .
\label{a3}
\eeq
To compute   (\ref{a2})   we write   the scalar field as $\phi(\x) = \phi_0  + \tilde{\phi}(\x)$, where $\phi_0$ is a  constant
and  $\tilde{\phi}$ is  an orthogonal  integrable function.  Similarly, we write $f(\x) = f_0 + \tilde{f}(\x)$ so that
(\ref{a2})  becomes
\beq
Z_{M}( f, \rho) =
 \int_{ \tilde{\phi}|_{ \partial {M}}  =  \tilde{f}   } d \phi_0   [ d \tilde{\phi}] \; \delta( \phi_0 - f_0)    \, e^{ - S_{M}(\tilde{\phi})  + i \phi_0 \sum_{j=1}^N q_j      }
 =   e^{ i f_0 \sum_{j=1}^N q_j  }    \int_{ \tilde{\phi}|_{ \partial {M}}  =  \tilde{f}   } [ d \tilde{\phi}] \;
     \, e^{ - S_{M}(\tilde{\phi})    } \, ,
\label{a2c}
\eeq
where $\delta( \phi_0 - f_0)$ implements the constraint $\phi(x) |_{\partial M} = f(x)$ between  the zero modes.
The second factor in (\ref{a2c})  can be computed explicitly obtaining
\beq
Z_{M}( f, \rho)  \propto   e^{ i f_0 \sum_{j=1}^N q_j  }   \;   e^{ - S_{M}(\phi_{\rm cl}) }   \, ,
\label{a4}
\eeq
where $\phi_{\rm cl}$ satisfies the  Poisson equation,
\beq
\partial_{x_i}^2 \phi_{\rm cl}( \x) = - \rho(\x), \qquad \phi_{\rm cl}( \x)|_{\x \in M}  =  \tilde{f}(\x) \, .
\label{a5}
\eeq
that is solved  by
\beq
\phi_{\rm cl}(\x) = -  \int_{M}  d^2 x' \,  G_M(\x, \x') \,  \rho( \x') -  \int_{\partial{M}} dx'_i  \,   \tilde{f}(\x') \,    \epsilon_{ij}  \partial_{x'_j}  G_M(\x, \x')      \, ,
\label{a6}
\eeq
where  $dx'_i$ is  the line element  along the curve $\partial M$ oriented anti-clockwise,  $\epsilon_{12} = - \epsilon_{21} = 1$,
and $G_M$ is the Green's function with Dirichlet BCs
\beq
\partial_{x_i}^2 G_M( \x, \x') = \delta^2(\x - \x'), \qquad G_M(\x, \x')=0, \quad
{\rm if} \; \x \;   {\rm or} \;  \x' \in  \partial M  \, .
\label{a7}
\eeq
From (\ref{a3}) we get
\barray
S_{M}( \phi_{\rm cl}) \equiv S_M(\tilde{f}, \rho)    &= &
  -  \frac{1}{8 \pi}  \int_{\partial M}  dx_i  \, \epsilon_{ij} \,   \phi_{\rm cl} \,  \partial_{x_j}  \phi_{\rm cl}
 - \frac{1}{8 \pi}  \int_{M}  d^2 x  \; \phi_{\rm cl}  \,  \rho \, ,
 \label{a8}
 \earray
where we have performed  a partial integration using  the Gauss  theorem,
\beq
\int_{M} d^2 x \;  \partial_i V_i = -   \int_{\partial M} dx_i \,  \epsilon_{ij} V_j , \quad V_i  = \phi_0 \partial_{x_i} \phi_{\rm cl}  \, .
\label{a9}
\eeq
Inserting  eq.(\ref{a6}) into (\ref{a8}) gives
\barray
S_{M}(\tilde{f}, \rho)  & = &  \frac{1}{8 \pi} \int_M   d^2 x  \int_M \, d^2 x' \, G_M( \x, \x') \rho(\x) \rho(\x')
\label{a10} \\
& + & \frac{1}{8 \pi} \int_{\partial M}    dx_i   \int_{\partial M}  \, d x'_k \,  \epsilon_{ij} \epsilon_{kl} \,  \tilde{f}(\x) \tilde{f}(\x') \,  \partial_{x_j} \partial_{x'_l}
G_M( \x, \x') \nonumber   \\
& + & \frac{1}{4 \pi} \int_{\partial M}    dx_i   \int_M \, d^2  x'  \,  \epsilon_{ij}  \tilde{f} (\x)   \rho(\x')  \partial_{x_j}  G_M( \x, \x')  \, ,
\nonumber
\earray
and then
\beq
Z_{M}( f, \rho)  \propto   e^{ i f_0 \sum_{j=1}^N q_j  }   \;   e^{ - S_M(\tilde{f}, \rho)   }   \, .
\label{a4b}
\eeq
This expression can  also be applied to the case when  $M$ is the sphere $S^2$.
Since $S^2$ has no boundary  the last two
terms of (\ref{a10}) are absent. The Green's function on $S^2$ is
\beq
G(z, \bar{z};  z', \bar{z}') = \frac{1}{4 \pi} \log |z - z'|^2  \, ,
\label{a11}
\eeq
where  $z = x_1 + i x_2$ and  $z' = x'_1 + i x'_2$.
In the rest of the SM we shall use the variable $z= x + i y$,
that  corresponds to   $t + i x$ in the main text.
We also have to integrate over $f_0$. The final result is
\beq
Z_{S^2} (\rho)  \propto  \delta (  \sum_{j=1}^N  q_j  )  \prod_{i >  j} |z_i - z_j|^{ 2  q_i q_j}  \, .
\label{a13}
\eeq
that does not vanish under  the neutrality condition
\beq
 \sum_{j=1}^N  q_j = 0  \,.
\label{a12bb}
\eeq
In the case discussed in the main text the charges are given by
\beq
q_j = q s_j,  \quad s_j = \pm 1  \, ,
\label{a12c}
\eeq
and  hence eq.(\ref{a12bb})  becomes  $\sum_{j=1}^N  s_j =0$.

\section{The MPS functional}

We shall use below the results obtained above  to construct the MPS functional.
First of all, we shall  find  the Green's function that solves eq.(\ref{a7}).
This can be done using  the Riemann's  mapping
theorem that asserts   the existence of a conformal map $g$ from $M$ to the upper-half plane $\Hmath$,
when $M$ is a simply connected region of the complex plane $\Cmath$
\beq
g: M \rightarrow \Hmath \,   , \qquad \Hmath = \{ z \in \Cmath, \;  {\rm Im} \; z \geq  0 \}  \, ,
\label{a14}
\eeq
such that the boundary of $M$,  is mapped into the real axis, that is $g : \partial M \rightarrow \Rmath$.
This map allows  us to construct $G_M$ from the Green's function $G_{\Hmath}$ in $\Hmath$, that is given by
\beq
G_{\Hmath} ( \zeta, \bar{\zeta};   \zeta', \bar{\zeta}')
= \frac{1}{4 \pi}  \log  \frac{ ( \zeta- \zeta') ( \bar{\zeta} - \bar{\zeta}')} { ( \zeta- \bar{\zeta}') ( \bar{\zeta} - {\zeta}')} \,,
\qquad  \zeta, \zeta'  \in \Hmath  \, .
\label{a15}
\eeq
  Notice that $G_{\Hmath}$ vanishes if  $\zeta$ or $\zeta'$ are real satisfying the Dirichlet BCs (\ref{a7}).
The Green's function $G_M$ can be obtained replacing $\zeta$ and $\zeta'$ by  $g(z)$ and $g(z')$ respectively,
\beq
G_M(z, \bar{z}; z', \bar{z}')
= \frac{1}{4 \pi}  \log  \frac{ ( g(z) - g(z')) ( \overline{g(z)} -  \overline{g(z')}  )} { ( g(z) -  \overline{g(z')} ( \overline{g(z)}   - g(z'))} \,,
\qquad  \; z, z' \in M  \, .
\label{a16}
\eeq
To construct the MPS  functional  we take the strip $M  =  \Rmath \times [0, \pi] $. The corresponding  conformal map (\ref{a14}) is given by
\beq
\zeta = e^z \, ,
\label{a17}
\eeq
that replaced into (\ref{a16}) gives
\beq
G_{M} (z, \bar{z}; z', \bar{z}')
= \frac{1}{4 \pi}  \log  \frac{ \sinh ( \frac{ z - z'}{2} ) \sinh ( \frac{  \bar{z}  - \bar{z}'}{2} ) } {  \sinh ( \frac{ z - \bar{z}'}{2} ) \sinh ( \frac{  \bar{z}  - {z}'}{2} )} \,,
\qquad  \; z, z' \in M  \, .
\label{a18}
\eeq

In order to prove the sewing and closing conditions,   given in eqs.(5a) and (5b) of the main text,
we shall  consider a generic strip $M  = \Rmath  \times \pi [a,b] \;(a < b)$, that can be mapped into $\Hmath$ by
the conformal map
\beq
g(z) = e^{ (z - i \pi a)/\Delta}, \qquad \Delta = b - a \, .
\label{a201}
\eeq
The associated Green's function is
\beq
G_{M} (z, \bar{z}; z', \bar{z}')
= \frac{1}{4 \pi}  \log  \frac{ \sinh ( \frac{ z - z'}{2 \Delta} ) \sinh ( \frac{  \bar{z}  - \bar{z}'}{2 \Delta} ) } {  \sinh ( \frac{ z - \bar{z}'}{2 \Delta} ) \sinh ( \frac{  \bar{z}  - {z}'}{2 \Delta} )} \,,
\qquad  \; z, z' \in M  \, .
\label{a202}
\eeq
Choosing  $z = x + i y, z' = x' + i y'$ we get
\barray
G_M(x,y; x',y') & = &  \frac{1}{4 \pi} \log  \frac{   \sinh  \frac{ x - x' + i (y - y')}{ 2 \Delta}  \sinh   \frac{ x - x' - i (y - y')}{ 2 \Delta}   }{
 \sinh  \frac{ x - x' - i (y + y') + 2 \pi i a }{ 2 \Delta}  \sinh   \frac{ x - x' + i (y + y') - 2 \pi i a}{ 2 \Delta} }   \, ,
 \label{a203} \\
\nonumber \\
\partial_y G_M(x,y; x',y') & = &
\frac{i}{8 \pi \Delta} \left[   \coth \frac{ x- x' + i (y - y') }{ 2 \Delta} -    \coth \frac{ x- x' -  i (y - y') }{ 2 \Delta}  \right.  \,
\nonumber  \\
& & +    \left.
\coth \frac{ x- x' -  i (y + y') + 2 \pi i a  }{ 2 \Delta} -    \coth \frac{ x- x' + i (y + y') - 2 \pi i a}{ 2 \Delta}   \right]  \, ,
\nonumber   \\
\partial_y  \partial_{y'} G_M(x,y; x',y') & = &
- \frac{1}{16 \pi  \Delta^2}   \left[  \frac{ 1}{ \sinh^2  \frac{ x - x' + i (y - y')}{ 2 \Delta} }   +  \frac{ 1}{ \sinh^2  \frac{ x - x' - i (y - y')}{ 2 \Delta}}
\right. \nonumber  \\
& & \left.   +   \frac{ 1}{ \sinh^2  \frac{ x - x' -  i (y + y') + 2 \pi i a}{ 2 \Delta} }   +  \frac{ 1}{ \sinh^2  \frac{ x - x' + i (y + y') - 2 \pi i a}{ 2 \Delta}} \right]
\, .
\nonumber
\earray

\subsection{The non chiral MPS functional}

The boundary $\partial M$ consists of the straight   lines in the plane  with $y=\pi a$ and $y = \pi b$.
We define the  real functions  (see fig. \ref{Tensors})
\beq
f_+(x) = \tilde{f}(x, \pi a), \quad f_-(x) =  \tilde{f}(x , \pi b) , \qquad x \in \Rmath \, .
\label{a21}
\eeq
The constant mode $f_0$ is common to both lines and will  be treated separately.
Its  contribution to the functional  is simply the phase factor in eq.(\ref{a4b}).
The functions $f_\pm(x)$ correspond to  $\alpha(t)$ and $\beta(t)$ in the main text.
The definitions (\ref{a21}) lead us to write  eq.(\ref{a10}) as
%%%%%
\barray
S_{M}[f_+, f_-, \rho]  &  = &  \frac{1}{8 \pi} \int_M   d^2 x  \int_M \, d^2 x' \, G_{M}( \x, \x') \rho(\x) \rho(\x')
\label{a22} \\
 & + & \frac{1}{8 \pi} \int_{\Rmath}    dx    \int_{\Rmath}  \, d x'  \,   ( f_+(x),  f_-(x))  \,
 \left( \begin{array}{cc}
  \partial_{y} \partial_{y'}  G_{M}(x, y; x' , y')|_{y_+, y_+'}    &
 -    \partial_{y} \partial_{y'}  G_{M}(x, y; x' , y')|_{ y_+, y'_-}  \\
-   \partial_{y} \partial_{y'}  G_{M}(x, y; x' , y')|_{y_-, y'_+} &
  \partial_{y} \partial_{y'}  G_{M}(x, y; x' , y')|_{y_-, y'_-} \\
  \end{array}
  \right)   \left( \begin{array}{l}
  f_+(x') \\
   f_-(x') \\
   \end{array}
   \right)
 \nonumber   \\
& + & \frac{1}{4 \pi} \int_{\Rmath}    dx    \int_M \, d^2  x'  \,  \rho(\x')   ( f_+(x),  f_-(x))
   \left(
\begin{array}{c}
  \partial_{y}  G_{M}(x, y; x' , y')|_{y_+} \\
-   \partial_{y}  G_{M}(x, y; x' , y')|_{y_-}\\
\end{array}
  \right)  \, ,
\nonumber
\earray
where
\beq
y_+ = \pi ( a + \varepsilon),  \quad y'_+ = \pi ( a + \varepsilon'),
\quad
y_- = \pi ( b - \varepsilon),  \quad y'_- = \pi ( b - \varepsilon'), \quad 0 < \varepsilon ,  \varepsilon' \ll 1 , \quad \varepsilon \neq \varepsilon'  \, ,
\label{a223}
\eeq
are a regularization of $y = \pi a, \pi b$. Eqs.(\ref{a203}) lead to
\barray
S_{M}[f_+, f_-, \rho] & = &  \frac{1}{8 \pi} \int_M   d^2 x  \int_M \, d^2 x' \, G_{M}( \x, \x') \rho(\x) \rho(\x')
\label{a23} \\
& - & \frac{1}{64 \pi^2} \int_{\Rmath}    dx    \int_{\Rmath}  \, d x'  \,   ( f_+(x),  f_-(x))  \,
 \left( \begin{array}{cc}
u_{+, \Delta}(x-x') &    u_{-, \Delta}(x-x') \\
u_{-, \Delta}(x-x') &    u_{+, \Delta}(x-x') \
  \end{array}
  \right)   \left( \begin{array}{l}
  f_+(x') \\
   f_-(x') \\
   \end{array}
   \right)
 \nonumber   \\
& + & \frac{i}{16 \pi^2} \int_{\Rmath}    dx    \int_M  \, d^2  x'  \,  \rho(\x')   ( f_+(x),  f_-(x))
  \left(
\begin{array}{r}
v_{+, \Delta,a}(x, \x') -  \overline{v_{+, \Delta,a}(x, \x') } \\
- v_{-,\Delta,a}(x, \x' ) + \overline{v_{-, \Delta,a}(x, \x')} \\
\end{array}
  \right) \, ,
\nonumber
\earray
with
\barray
u_{+, \Delta}(x-x') & =  &  \frac{1}{\Delta^2} \left(  \frac{1}{ \sinh^2  \frac{x - x' + i \epsilon}{2 \Delta} }   +
 \frac{1}{ \sinh^2  \frac{x - x' -  i \epsilon}{2 \Delta} }
   \right)  \; , \quad    \label{a24}  \\
u_{-, \Delta}(x-x') &  =   &   \frac{2}{ \Delta^2    \cosh^2  \frac{x - x' }{2 \Delta} } \, ,
\label{a24b}   \\
v_{+, \Delta,a}(x, \x') & =  &  \frac{1}{\Delta}  \coth \frac{ x - x' - i y' + i \pi a }{2 \Delta}  \;  ,
\quad v_{-, \Delta,a} (x, \x')   = \frac{1}{\Delta}
  \tanh \frac{ x - x' - i y' + i \pi a }{2 \Delta} \;  .
\label{a24c}
\earray
In equations \eqref{a24b} and \eqref{a24c} we have taken the limit $\varepsilon, \varepsilon' \rightarrow 0$
that gives ordinary functions, while in \eqref{a24} we have replaced $\varepsilon \pm \varepsilon'$ by $\epsilon$
that in the limit $\epsilon \rightarrow 0$ becomes  a generalized function, namely  a distribution.
To discuss this issue in more detail we define the function
\beq
f_\epsilon(x) = \frac{1}{ \sinh^2 ( x + i \epsilon)} +  \frac{1}{ \sinh^2 ( x - i \epsilon)} \, ,
\label{dis1}
\eeq
that is obtained from   \eqref{a24} after rescaling the  variables.
Let us also define the function
\beq
g_\epsilon(x) = 2 \left(  \frac{1}{ \sinh^2 x} - \frac{1}{x^2} \right)   + \frac{1}{( x + i \epsilon)^2} + \frac{1}{(x - i \epsilon)^2}
\label{dis2}
\eeq
whose first term is regular at $x=0$. We aim at showing  that $f_\epsilon(x) \rightarrow g_\epsilon(x)$ in the limit
$\epsilon \rightarrow 0$.
Expanding the difference between \eqref{dis1} and \eqref{dis2} around $\epsilon=0$
yields
\beq
f_\epsilon(x)  - g_\epsilon(x) = \left( \frac{6}{x^4} - \frac{ 2 ( 2 + \cosh( 2 x))}{\sinh^4 x} \right)  \epsilon^2 + O(\epsilon^4) \, .
\label{dis3}
\eeq
The term proportional to $\epsilon^2$, has the series  expansion $- \frac{2}{15} + O(x^2)$ around $x=0$,
so that one  can take safely the limit $\epsilon \rightarrow 0$ obtaining
\beq
\lim_{\epsilon \rightarrow 0} (f_\epsilon(x) - g_\epsilon(x) ) = 0 , \quad \forall x \in \Rmath \, .
\label{dis4}
\eeq
The terms depending of $\epsilon$  in \eqref{dis2} can be expressed
using the principal value distribution,
\beq
P \left( \frac{1}{x} \right) = \lim_{\epsilon \rightarrow 0} \frac{1}{2} \left( \frac{1}{x + i \epsilon} + \frac{1}{x - i \epsilon} \right) \, ,
\label{dis5}
\eeq
whose derivative respect to $x$ is
\beq
P' \left( \frac{1}{x} \right) = -  \lim_{\epsilon \rightarrow 0} \frac{1}{2} \left( \frac{1}{(x + i \epsilon)^2} + \frac{1}{(x - i \epsilon)^2} \right) \ ,
\label{dis6}
\eeq
that together with \eqref{dis2} and \eqref{dis4} yields
\beq
\lim_{\epsilon \rightarrow 0} f_\epsilon(x) = 2 \left(  \frac{1}{ \sinh^2 x} - \frac{1}{x^2}  - P' \left( \frac{1}{x} \right)  \right)  \, .
\label{dis7}
\eeq
Rescaling the  variables leads finally to
\beq
\lim_{\epsilon \rightarrow 0} u_{+, \Delta}(x-x') = \frac{2}{\Delta^2}
\left( \frac{1}{ \sinh^2  \frac{x - x' }{2 \Delta} }  - \left(  \frac{2 \Delta}{x - x'} \right)^2 \right)  - 8 \, P' \left( \frac{1}{x - x'} \right) \, ,
\label{dis8}
\eeq
that coincides with  the Eq.(14)  given in the main text with $\Delta = L$ and $x- x' \rightarrow t$.

\subsection{The chiral  MPS functional}

Eq.(\ref{a23})  is the  basis of  our  proposal for  a chiral version of the MPS functional.
It is obtained by  keeping the terms that depend  exclusively  on  $z$ or  $z'$
 in the Green's function
and in the piece  proportional to $\rho(\x')$.
The terms quadratic
in  $f_\pm$  do  not possess a chiral/antichiral factorization and stay the same.
The chiral functional is defined by  truncating  $S_M$ to
\barray
R_{M}[f_+, f_-, \rho] & = &  \frac{1}{32 \pi^2  } \int_M   d^2 x  \int_M \, d^2 x' \, \rho(\x) \rho(\x')  \log ( \mu  \sinh  \frac{ x - x' + i (y - y')}{2 \Delta} )
\label{a25} \\
& - & \frac{1}{64 \pi^2} \int_{\Rmath}    dx    \int_{\Rmath}  \, d x'  \,   ( f_+(x),  f_-(x))  \,
 \left( \begin{array}{cc}
u_{+, \Delta}(x-x') &    u_{-, \Delta}(x-x') \\
u_{-, \Delta}(x-x') &    u_{+, \Delta}(x-x') \
  \end{array}
  \right)   \left( \begin{array}{l}
  f_+(x') \\
   f_-(x') \\
   \end{array}
   \right)
 \nonumber   \\
& + & \frac{i}{16 \pi^2} \int_{\Rmath}    dx    \int_M  \, d^2  x'  \,  \rho(\x')   ( f_+(x),  f_-(x))
  \left(
\begin{array}{r}
v_{+, \Delta,a}(x, \x') \\
- v_{-, \Delta,a} (x, \x' ) \\
\end{array}
  \right) \, .
\nonumber
\earray
We have introduced a constant $\mu$ whose value will be fixed  later on.
Replacing  the charge density (\ref{a12}) in (\ref{a25}) gives
\barray
R_{M}[f_+, f_-, \{ q_j , z_j \}_{j=1}^N] & = &  -  \sum_{N \geq j > k \geq 1}   q_j q_k  \log ( \mu   \sinh \frac{ z_j - z_k}{2 \Delta}  )
\label{a26} \\
& - & \frac{1}{64 \pi^2} \int_{\Rmath}    dx    \int_{\Rmath}  \, d x'  \,   ( f_+(x),  f_-(x))  \,
 \left( \begin{array}{cc}
u_{+, \Delta}(x-x') &    u_{-, \Delta}(x-x') \\
u_{-, \Delta}(x-x') &    u_{+, \Delta}(x-x') \
  \end{array}
  \right)   \left( \begin{array}{l}
  f_+(x') \\
   f_-(x') \\
   \end{array}
   \right)
 \nonumber   \\
& -  & \frac{1}{4 \pi }  \sum_{j=1}^N  \int_{\Rmath}    dx  \;  q_j   ( f_+(x),  f_-(x))
  \left(
\begin{array}{r}
v_{+, \Delta,a}(x, z_j) \\
- v_{-, \Delta,a}(x, z_j ) \\
\end{array}
  \right) \, ,
\nonumber
\earray
 where we  have eliminated  the divergent  terms arising when   $j=k$ in the sum over the charges.
 The charges are located in the strip $M = \Rmath \times \pi [a, b]$, that is
 \beq
\pi a <  {\rm Im} \; z_j  < \pi b , \quad  \; j=1, \dots, N \, .
\label{a261}
\eeq

\begin{figure}[h!]
\includegraphics[width=110mm]{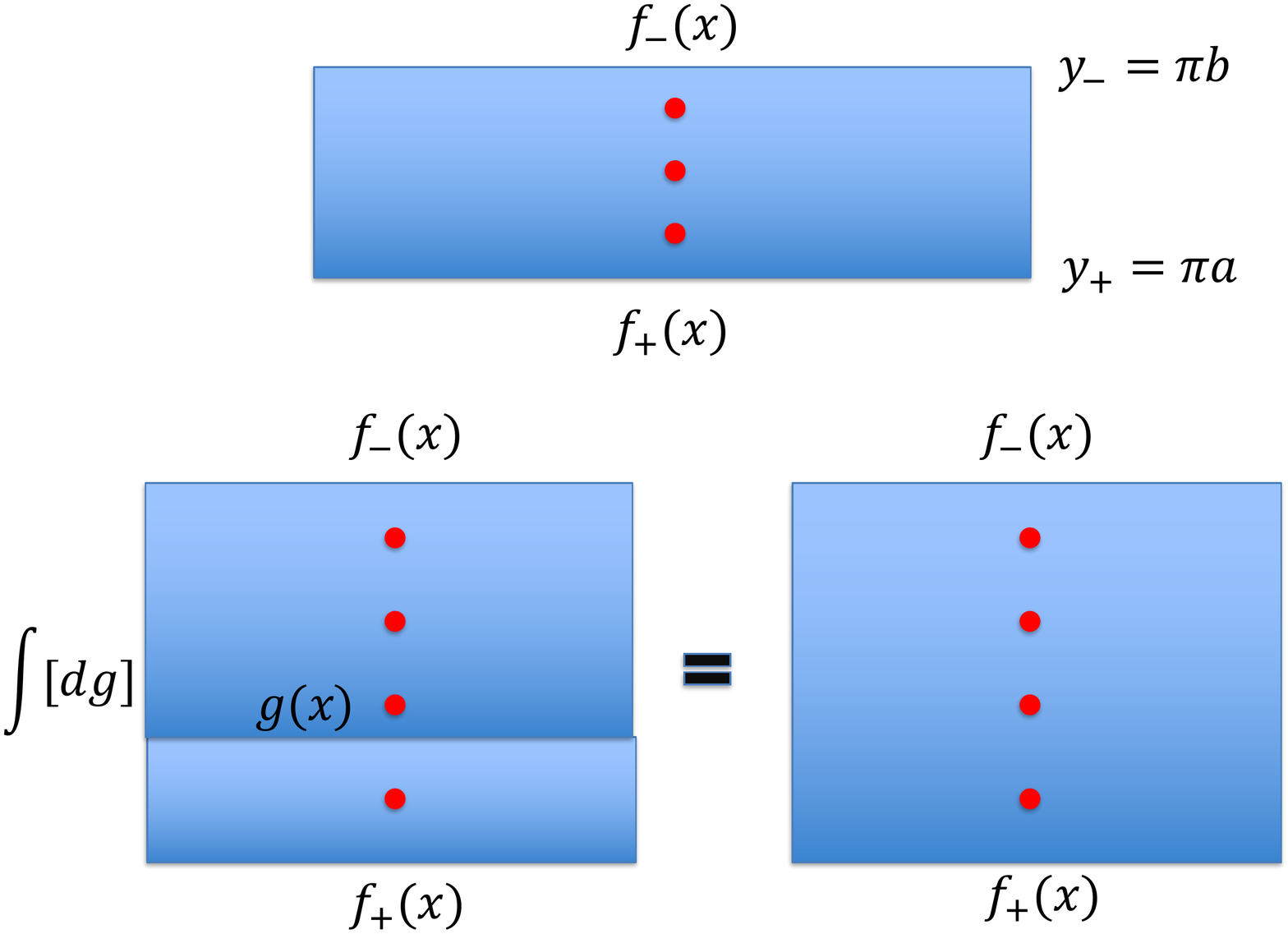}
\caption{Top: graphical representation of the  MPS functional (\ref{a27}) for $N=3$.
Bottom:  sewing condition (\ref{a28})}
\label{Tensors}
\end{figure}

We shall define the functional
\beq
A_{M}[f_+, f_-, \{ q_j , z_j \}_{j=1}^N] =    {\rm exp} \left( - R_{M}[f_+, f_-, \{ q_j , z_j \}_{j=1}^N] \right)  \, ,
\label{a27}
\eeq
that is represented in fig.\ref{Tensors}.
Observe that for  $N=1$  the
first term in eq.(\ref{a26}) does not appear. $A_{M}[f_+, g, \{ q , z \}] $  is the basic
building block   to construct  the functionals with  $N>1$. This is a consequence of the
sewing condition  illustrated in fig.\ref{Tensors},
\beq
\int [d g]
A_{M_1}[f_+, g, \{ q_0 , z_0 \}]   \;  A_{M_L}[g, f_-, \{ q_j , z_j \}_{j=1}^L]    \propto  A_{M_1 \cup M_L}[f_+, f_-, \{ q_j , z_j \}_{j=0}^L]    \, ,
\label{a28}
\eeq
 where $M_1 = \Rmath \times[- \pi , 0]$ and $M_L = \Rmath \times [0, \pi L]$ are two strips with
 a common boundary $M_1 \cap M_L = \Rmath \times \{ 0 \}$.
 The coordinates of the charges belong to the corresponding intervals, that is
$z_0 \in M_1$, $z_{j=1, \dots, L} \in M_L$. Equation (\ref{a28}) can be interpreted  geometrically as the sewing of the
strips  $M_1$ and $M_L$ along $M_1 \cap M_L$ to produce the strip $M_{L+1} \equiv M_1 \cup M_L = \Rmath \times [- \pi, \pi L]$.

The functional  for $M_L$ is  given by (\ref{a26}) and (\ref{a27}) with $N=L$ and $\Delta = L$, that is
\barray
A_{M_L}[f_+, f_-, \{ q_j , z_j \}_{j=1}^L] & = & {\rm exp} ( -  R_{M_L}[f_+, f_-, \{ q_j , z_j \}_{j=1}^L] ) \, , \label{a29}  \\
R_{M_L}[f_+, f_-, \{ q_j , z_j \}_{j=1}^L] & = &  -  \sum_{L \geq j > k \geq 1}   q_j q_k  \log  ( \mu  \sinh \frac{ z_j - z_k}{2L}  )
\nonumber  \\
& - & \frac{1}{64 \pi^2} \int_{\Rmath}    dx    \int_{\Rmath}  \, d x'  \,   ( f_+(x),  f_-(x))  \,
 \left( \begin{array}{cc}
u_{+, L}(x-x') &    u_{-,L}(x-x') \\
u_{-, L}(x-x') &    u_{+,L}(x-x') \
  \end{array}
  \right)   \left( \begin{array}{l}
  f_+(x') \\
   f_-(x') \\
   \end{array}
   \right)
 \nonumber   \\
& -  & \frac{1}{4 \pi }  \sum_{j=1}^L   \int_{\Rmath}    dx  \;  q_j   ( f_+(x),  f_-(x))
  \left(
\begin{array}{r}
v_{+, L}(x, z_j) \\
- v_{-, L}(x, z_j ) \\
\end{array}
  \right) \, .
\nonumber
\earray
If $L=1$ the log term does not appear.
Similarly,  the functional  for $M_1$ is given by
\barray
A_{M_1}[f_+, f_-, \{ q_0 , z_0 \} ] & = & {\rm exp} ( -  R_{M_1}[f_+, f_-, \{ q_0 , z_0 \}] ) \, , \label{a30}  \\
R_{M_1}[f_+, f_-, \{ q_0 , z_0 \}] &=   & -  \frac{1}{64 \pi^2} \int_{\Rmath}    dx    \int_{\Rmath}  \, d x'  \,   ( f_+(x),  f_-(x))  \,
 \left( \begin{array}{cc}
u_{+, 1}(x-x') &    u_{-,1}(x-x') \\
u_{-,1}(x-x') &    u_{+,1}(x-x') \
  \end{array}
  \right)   \left( \begin{array}{l}
  f_+(x') \\
   f_-(x') \\
   \end{array}
   \right)
 \nonumber   \\
& -  & \frac{1}{4 \pi }    \int_{\Rmath}    dx  \;  q_0  \;   ( f_+(x),  f_-(x))
  \left(
\begin{array}{r}
v_{-, 1}(x, z_0) \\
- v_{+, 1}(x, z_0 ) \\
\end{array}
  \right) \, ,
\nonumber
\earray
where from (\ref{a24}) one has
\barray
u_{+, L}(x-x') & =  &  \frac{1}{L^2} \left(  \frac{1}{ \sinh^2  \frac{x - x' + i \epsilon}{2 L} }   +
 \frac{1}{ \sinh^2  \frac{x - x' -  i \epsilon}{2 L} }
   \right)  \; , \quad    \label{a31}  \\
u_{-, L}(x-x') &  =   &   \frac{2}{ L^2    \cosh^2  \frac{x - x' }{2L} } \, ,
 \nonumber  \\
v_{+, L}(x, z') & \equiv   &  v_{+, L,0}(x, z')  =  \frac{1}{L}  \coth \frac{ x - z' }{2 L}  \; , \nonumber \\
\quad v_{-, L} (x, z')  &  \equiv &   v_{-, L,0}(x, z') =  \frac{1}{L}
  \tanh \frac{ x - z'  }{2 L} \;  .
  \nonumber
\earray
In the last term of eq.(\ref{a30}) we used that
\beq
v_{\pm, 1,-1}(x, z')  = v_{\mp, 1, 0}(x, z')  = v_{\mp}(x, z') \, .
\label{a311}
\eeq

\subsection{The chiral MPS functional  in momentum space}

To perform   the path integral (\ref{a28}), we exploit the translation invariance of $u_{\pm , L}(x-x')$
working in momentum space. First of all,  we define the Fourier transform $\hat{f}_\pm(k)$  of the integrable functions $f_\pm(x)$,
\beq
f_{\pm}(x) = \int_\Rmath dk \;  e^{ i k x} \hat{f}_\pm(k)  \, .
\label{a32}
\eeq
The reality of $f_\pm(x)$ implies that $\hat{f}_\pm(-k) = \hat{f}_\pm^*(k)$.
The Fourier transform of the functions (\ref{a31}) is
\barray
\int_\R d x \, e^{ i k x} \, u_{+,L}(x)  & = &
 - 8 \pi k \coth( \pi k L) ,  \label{a33} \\
\int_\R d x \, e^{ i k x} \, u_{-, L}(x)  &  = &  8 \pi k/ \sinh(\pi k L)  \, ,  \nonumber   \\
\int_R dx \,  e^{ i k x} \, v_{+,L}(x, z')  & = &  2  \pi i   \, e^{ i k z'} e^{ \pi k L}/\sinh(\pi k L) ,
\qquad 0 < {\rm Im} \, z'  < \pi L  \, ,   \nonumber   \\
\int_R dx \,  e^{ i k x} \,  v_{+,L}(x,z')  & = &   2 \pi i   \, e^{ i k z'} e^{ - \pi k L}/\sinh(\pi k L)  ,
\quad - \pi  < {\rm Im} \, z'  < 0  \, ,   \nonumber  \\
\int_R \,   e^{ i k x} \, v_{-,L}(x, z') & = &   2 \pi i   \, e^{ i k z'}/\sinh(\pi k L)  ,
\quad - \pi L  < {\rm Im} \, z'< \pi L  \, .   \nonumber
 \earray
The functionals (\ref{a29}) and (\ref{a30}) become in momentum space
\barray
R_{M_L}[f_+, f_-, \{ q_j , z_j \}_{j=1}^L] & = &  -  \sum_{L \geq j > k \geq 1}   q_j q_k  \log  ( \mu \sinh \frac{ z_j - z_k}{2 L} )
\label{a35} \\
&+  & \frac{1}{2} \int_{0}^\infty     dk \;    ( \hat{f}_+(k),  \hat{f}_-(k))  \,
 \left( \begin{array}{cc}
\omega_{+,L}(k) &    \omega_{-,L}(k) \\
\omega_{-,L}(k) &    \omega_{+,L}(k) \
  \end{array}
  \right)   \left( \begin{array}{l}
  \hat{f}_+^*(k) \\
   \hat{f}_-^*(k) \\
   \end{array}
   \right)
 \nonumber   \\
& -  & \frac{i}{2 }  \sum_{j=1}^L   q_j  \int_{\Rmath}    dk  \;    \frac{ e^{ i k z_j}}{ \sinh( \pi k L)}   ( e^{ \pi k L}  \hat{f}_+(k)-    \hat{f}_-(k))   \, ,
\nonumber
\earray
and
\barray
R_{M_1}[f_+, f_-, \{ q_0 , z_0 \}] & = &
 \frac{1}{2} \int_{0}^\infty     dk \;    ( \hat{f}_+(k),  \hat{f}_-(k))  \,
 \left( \begin{array}{cc}
\omega_{+,1}(k) &    \omega_{-,1}(k) \\
\omega_{-,1}(k) &    \omega_{+,1}(k) \
  \end{array}
  \right)   \left( \begin{array}{l}
  \hat{f}_+^*(k) \\
   \hat{f}_-^*(k) \\
   \end{array}
   \right)
 \label{a34}   \\
& -  & \frac{i q_0}{2 }   \int_{\Rmath}    dk  \;  \frac{ e^{ i k z_0}}{ \sinh( \pi k)}   ( \hat{f}_+(k)- e^{ - \pi k}   \hat{f}_-(k))   \, ,
\nonumber
\earray
where
 \beq
 \omega_{+,L}(k) = k  \coth(\pi k L), \qquad
  \omega_{-,L}(k) = - k/\sinh(\pi k L)  \, .
 \label{a36}
 \eeq

 \subsection{The sewing condition}

The LHS of (\ref{a28}) is given by
\beq
\int [d g]
A_{M_1}[f_+, g, \{ q_0 , z_0 \}]   \;  A_{M_L}[g, f_-, \{ q_j , z_j \}_{j=1}^L]    =
 \int [d g]  {\rm exp} \left( -
R_{M_1}[f_+, g, \{ q_0 , z_0 \}]   - R_{M_L}[g, f_-, \{ q_j , z_j \}_{j=1}^L]   \right)  \, .
\label{a37}
\eeq
The exponent of the integrand  is the sum
\beq
R_{M_1} + R_{M_L} =  \circled{1} +   \circled{2} +  \circled{3} +   \circled{4} +  \circled{5}
\label{a38}
\eeq
where
\barray
\circled{1}  & = & -  \sum_{L\geq  j > k \geq  1}   q_j q_k  \log ( \mu  \sinh \frac{ z_j - z_k}{2 L} )  \, ,  \label{a39} \\
\circled{2}  & = &  \frac{1}{2} \int_{0}^\infty     dk \;    ( \hat{g}(k),  \hat{f}_-(k))  \,
 \left( \begin{array}{cc}
\omega_{+,L}(k) &    \omega_{-,L}(k) \\
\omega_{-,L}(k) &    \omega_{+,L}(k) \
  \end{array}
  \right)   \left( \begin{array}{l}
  \hat{g}^*(k) \\
   \hat{f}_-^*(k) \\
   \end{array}
   \right)    \, ,   \nonumber  \\
\circled{3}  & = &   - \frac{i}{2 }  \sum_{j=1}^L  q_j  \int_{\Rmath}
dk  \;    \frac{ e^{ i k z_j}}{ \sinh( \pi k L)}   ( e^{ \pi k L}  \hat{g}(k)-    \hat{f}_-(k))   \, ,     \nonumber  \\
\circled{4}  & = &    \frac{1}{2} \int_{0}^\infty     dk \;    ( \hat{f}_+(k),  \hat{g}(k))  \,
 \left( \begin{array}{cc}
\omega_{+,1}(k) &    \omega_{-,1}(k) \\
\omega_{-,1}(k) &    \omega_{+,1}(k) \
  \end{array}
  \right)   \left( \begin{array}{l}
  \hat{f}_+^*(k) \\
   \hat{g}^*(k) \\
   \end{array}
   \right)  \,   ,    \nonumber  \\
\circled{5}  & = & - \frac{i q_0}{2 }   \int_{\Rmath}    dk  \;  \frac{ e^{ i k z_0}}{ \sinh( \pi k)}   ( \hat{f}_+(k)- e^{ - \pi k}   \hat{g}(k))  \, .    \nonumber
\earray
Consider the partial sums
\barray
\circled{2} + \circled{4}  & = &  \circled{7} + \circled{8}
\label{a391}  \\
\circled{3} + \circled{5}  & = &  \circled{9} + \circled{10}
\nonumber
\earray
where

\barray
\circled{7} & =  &
 \frac{1}{2} \int_0^\infty dk \;
\left[ \hat{g}(k) \hat{g}^*(k)   (  \omega_{+, L}(k) + \omega_{+,1}(k))    \right.   \label{a40} \\
& &  \left.   +  \hat{g}(k) (   \hat{f}_-^*(k)  \omega_{-,L}(k)     +    \hat{f}_+^*(k)  \omega_{-,1}(k)   ) +
 g^*(k)   (   \hat{f}_-(k)  \omega_{-,L}(k)     +    \hat{f}_+(k)  \omega_{-,1}(k)   )
    \right]  \, ,
\nonumber  \\
\circled{8} & = &  \frac{1}{2} \int_0^\infty dk \;
\left[   \hat{f}_+(k)    \hat{f}_+^*(k)  \omega_{+,1}(k)    +   \hat{f}_-(k)    \hat{f}_-^*(k)  \omega_{+,L}(k)   \right]  \, ,  \nonumber  \\
\circled{9} & = &- \frac{i}{2} \int_0^\infty  dk  \;
 \left[  q_0   \frac{   e^{ i k z_0} \hat{f}_+(k) - e^{- i k z_0} \hat{f}_+^*(k)  }{ \sinh (\pi k)}
  - \sum_{j=1}^L q_j   \frac{   e^{ i k z_j} \hat{f}_-(k) -  e^{ - i k z_j} \hat{f}_-^*(k)   }{ \sinh( \pi k L)}    \right]  \, ,  \label{a41} \\
\circled{10} & = &  - \frac{i}{2} \int_0^\infty  dk  \left[
\hat{g}(k) \left(    \sum_{j=1}^L   q_j  \frac{ e^{ i k z_j} e^{ \pi k L} }{ \sinh (\pi k L)} - q_0   \frac{ e^{i k z_0} e^{ - \pi k} }{ \sinh (\pi k)}  \right)
- \hat{g}^*(k) \left(    \sum_{j=1}^L   q_j  \frac{ e^{ - i k z_j} e^{ -  \pi k L} }{ \sinh (\pi k L)} - q_0   \frac{ e^{- i k z_0} e^{  \pi k} }{ \sinh (\pi k)}  \right)
\right]  \, ,
\nonumber
\earray
such that
\beq
R_{M_1} + R_{M_L} =
 \circled{1} +   \circled{7} +  \circled{8} +   \circled{9} +  \circled{10}
\label{a411}
\eeq
The  terms depending  on  $g(k)$  are
\barray
  \circled{7} +  \circled{10}
 & = &
 \int_0^\infty dk \;  \left[
 \hat{g}(k) \hat{g}^*(k)   \Omega(k)   +   \hat{g}(k)   \alpha(k) +   \hat{g}^*(k)   \beta(k) \right]  \, ,
\label{a42}
 \earray
where
\barray
\Omega(k) & = & \frac{1}{2} ( \omega_{+,L}(k) + \omega_{+,1} (k)) = \frac{k}{2}  \frac{ \sinh( \pi k (L+1)}{ \sinh( \pi k L) \sinh(\pi k )} ,
\label{a43} \\
\alpha(k) & = & \frac{1}{2} (   \hat{f}_-^*(k)  \omega_{-, L}(k)   +  \hat{f}_+^*(k)  \omega_{-,1}(k)  ) - \frac{i}{2 }
 \left ( \sum_{j=1}^L q_j  \frac{ e^{ i k z_j} e^{ \pi k L} }{ \sinh( \pi k L)  }  - q_0  \frac{ e^{i k z_0} e^{- \pi k}}{ \sinh( \pi k)}  \right)  \, ,  \nonumber  \\
 \beta(k) & = & \frac{1}{2} (   \hat{f}_-(k)  \omega_{-, L}(k)   +  \hat{f}_+(k)  \omega_{-,1}(k)  ) + \frac{i}{2 }
 \left ( \sum_{j=1}^L q_j  \frac{ e^{ - i k z_j} e^{ - \pi k L} }{ \sinh( \pi k L)  }  - q_0  \frac{ e^{- i k z_0} e^{ \pi k}}{ \sinh( \pi k)}  \right)  \, .   \nonumber
 \earray
Now, we   perform the functional integral
\barray
\int \prod_{k>0} d \hat{g}(k) d \hat{g}^*(k) {\rm exp} \left[  -
 \int_0^\infty dk \;  \left(
 \hat{g}(k) \hat{g}^*(k)   \Omega(k)   +   \hat{g}(k)   \alpha(k) +   \hat{g}^*(k)   \beta(k) \right)
 \right]   \propto {\rm exp} \left(   \int_0^\infty  dk \frac{  \alpha(k) \beta(k)   }{  \Omega(k)}   \right)  \, ,
 \label{a44}
  \earray
where we  used
\beq
\int   dz \, d \bar{z}  \, e^{ - \omega |z|^2 +   a z +   b z^*} = \frac{ \pi}{\omega} e^{ a b/\omega} \, .
\label{a45}
\eeq
Upon  integration, the  LHS of (\ref{a37})  can be written as  $e^{ - \tilde{R}_{M_1 \cup M_L}}$ with
\beq
\tilde{R}_{M_1 \cup M_L} =
\circled{1} +   \circled{8} +  \circled{9}
 -   \int_0^\infty  dk \frac{  \alpha(k) \beta(k)   }{  \Omega(k)}  \, .
 \label{a451}
\eeq
Our  goal  is to show  that $e^{ - \tilde{R}_{M_1 \cup M_L}}$ coincides with $e^{ - {R}_{M_1 \cup M_L}}$.
The last  integral in (\ref{a451})  splits as
\beq
 -   \int_0^\infty  dk \frac{  \alpha(k) \beta(k)   }{  \Omega(k)}  =
\circled{11} +   \circled{12} +  \circled{13}
\label{a452}
\eeq
 where
\barray
\circled{11} & = &  -    \int_0^\infty  \frac{dk}{ 4 \Omega(k)}
 \left ( \sum_{j=1}^L q_j  \frac{ e^{ i k z_j} e^{ \pi k L} }{ \sinh( \pi k L)  }  - q_0  \frac{ e^{i k z_0} e^{- \pi k}}{ \sinh( \pi k)}  \right)
  \left ( \sum_{l=1}^L q_l  \frac{ e^{ - i k z_l} e^{ - \pi k L} }{ \sinh( \pi k L)  }  - q_0  \frac{ e^{- i k z_0} e^{ \pi k}}{ \sinh( \pi k)}  \right)
\label{a46} \\
& = & -    \int_0^\infty  \frac{dk}{ 4 \Omega(k)}   \left(
\sum_{L \geq j, l \geq 1} q_j q_l \frac{ e^{ i k (z_j - z_l)}}{ \sinh^2( \pi k L)}  + \frac{ q_0^2}{ \sinh^2(\pi k)}
- \sum_{j=1}^L q_j q_0  \frac{ e^{ i k (z_j - z_0)} e^{ \pi k (L+1)} + e^{ - i k (z_j - z_0)} e^{-  \pi k (L+1)} }{ \sinh(\pi k L) \sinh(\pi k)}
 \right)   \, ,
\nonumber   \\
\circled{12} & = &     -  \int_0^\infty  \frac{dk}{ 4 \Omega}
\left(  \hat{f}_-(k) \omega_{-,L}(k)  + \hat{f}_+(k) \omega_{-,1}(k) \right)
\left(  \hat{f}_-^*(k) \omega_{-,L}(k)  + \hat{f}_+^*(k) \omega_{-,1}(k)  \right) \, ,
 \nonumber  \\
\circled{13} & = &     i   \int_0^\infty  \frac{dk}{ 4 \Omega}
\left[ ( \hat{f}_+(k) \omega_{-,1}(k)  +  \hat{f}_-(k) \omega_{-,L}(k)  )
 \left( \sum_{j=1}^L q_j  \frac{ e^{ i k z_j} e^{ \pi k L} }{ \sinh( \pi k L)  }  - q_0  \frac{ e^{i k z_0} e^{- \pi k}}{ \sinh( \pi k)}  \right)
 \right.    \nonumber   \\
  & &  \qquad \qquad  \left. -  ( \hat{f}_+^*(k) \omega_{-,1}(k)  +  \hat{f}_-^*(k) \omega_{-,L}(k)  )
 \left( \sum_{j=1}^L q_j  \frac{ e^{-  i k z_j} e^{ -  \pi k L} }{ \sinh( \pi k L)  }  - q_0  \frac{ e^{- i k z_0} e^{ \pi k}}{ \sinh( \pi k)}  \right)
 \right]  \, .
 \earray
 The integrand of  \circled{11} behaves as $1/k^2$ for $k \sim 0$.
To obtain a finite value we use the following regularization method.
We  first write \circled{11}  as
\beq
\circled{11} = - \left[  \sum_{ L \geq j > l \geq 1} q_j q_l  \, I_{1,L}(z_j - z_l) - \sum_{j=1}^L q_j q_0 \, I_{2,L}(z_j - z_0)
+ \sum_{j=1}^L q_j^2  \, I_{3,L} + q_0^2 \, I_{4,L}\right] \, ,
\label{a47}
\eeq
with
\barray
I_{1,L}(z) & = &   \int_0^\infty  \frac{dk}{ 4 \Omega(k)}    \frac{ e^{ i k z} + e^{ - i k z}}{ \sinh^2( \pi k L)}  =
 \int_0^\infty  \frac{dk}{  k}    \frac{ \sinh( \pi k) \cos( k z )}{ \sinh( \pi k L) \sinh( \pi k (L+1))}  \, ,
\label{a48} \\
I_{2,L}(z) & = &   \int_0^\infty  \frac{dk}{ 4 \Omega(k)}    \frac{ e^{ i k z + \pi k (L+1) }  + e^{ -  i k z  - \pi k (L+1)}}{ \sinh( \pi k L) \sin (\pi k)}  =
 \int_0^\infty  \frac{dk}{  k}    \frac{ \cosh( i k z + \pi k (L+1) )}{  \sinh( \pi k (L+1))}  \, ,
 \nonumber  \\
 I_{3,L}(z) & = &   \int_0^\infty  \frac{dk}{ 4 \Omega(k)}    \frac{1}{ \sinh^2( \pi k L)}  =
 \int_0^\infty  \frac{dk}{  2 k}    \frac{ \sinh( \pi k) }{ \sinh( \pi k L) \sinh( \pi k (L+1))}  \, ,
 \nonumber  \\
 I_{4,L}(z) & = &   \int_0^\infty  \frac{dk}{ 4 \Omega(k)}    \frac{1}{ \sinh^2( \pi k L)}  =
 \int_0^\infty  \frac{dk}{  2 k}    \frac{ \sinh( \pi k L) }{ \sinh( \pi k ) \sinh( \pi k (L+1))}  \, .
 \nonumber
\earray
The integrands of these function is even, so one can replace the integration interval from $(0, \infty)$
to $ \Rmath$. Next,  we replace  $\Rmath$ by a contour  ${\cal R}_\epsilon $  in the complex plane
that  runs along the negative real axis until  the point $(- \varepsilon, 0)$, encircles the origin
clockwise until the point $(0, \epsilon)$ and continues along the positive real axis,
\beq
{\cal R}_\epsilon = (- \infty, \epsilon) \cup \{ \epsilon \; e^{ i \theta} |   \;  \theta \in (\pi, 0) \} \cup (\epsilon, \infty), \qquad
0 \leq \epsilon \ll 1 \, .
\label{a49}
\eeq
The regularized integrals (\ref{a48}) are  given by
\barray
I_{1,L}^{\epsilon}(z)  & = &
 \int_{{\cal R}_\epsilon}   \frac{dk}{ 2  k}    \frac{ \sinh( \pi k) \cos( k z )}{ \sinh( \pi k L) \sinh( \pi k (L+1))} ,
 \qquad  0 < | {\rm Im} (z)  | < \pi L
\label{a50} \\
I_{2,L}^\epsilon(z) & = &
 \int_{{\cal R}_\epsilon}   \frac{dk}{ 2  k}    \frac{ \cosh( i k z + \pi k (L+1) )}{  \sinh( \pi k (L+1))} ,
\qquad   \qquad 0 <   {\rm Im} (z)    < \pi (L + 1)
 \label{a501}  \\
I_{3,L}^\epsilon & = &
 \int_{{\cal R}_\epsilon}   \frac{dk}{ 4  k}    \frac{ \sinh( \pi k) }{ \sinh( \pi k L) \sinh( \pi k (L+1))}  ,
 \label{a502}  \\
I_{4,L}^\epsilon & = &
 \int_{{\cal R}_\epsilon}   \frac{dk}{ 4  k}    \frac{ \sinh( \pi k L) }{ \sinh( \pi k ) \sinh( \pi k (L+1))},
 \label{a503}
\earray
where we  included the ranges  of the variable $z$ that come from their relation to $z_j$ and $z_0$.

The simplest integral is
\beq
I_{3,1}^\epsilon  =  \int_{{\cal R}_\epsilon}   \frac{dk}{ 4  k}    \frac{ 1}{ \sinh( 2 \pi k )}  = - \frac{1}{2} \log 2 \, ,
\label{a51}
\eeq
that can be computed  using residue calculus.  Similarly
\beq
I_{3,L}^\epsilon  =   - \frac{1}{2} \log  \frac{L+1}{L} , \qquad  I_{4,L}^\epsilon  =   - \frac{1}{2} \log  (L+1)   \, .
\label{a52}
\eeq
To compute  $I_{2,L}^\epsilon(z)$ by residue calculus we close the contour ${\cal R}_\epsilon$
on a half circle of radius $R$  in the upper half plane, and  take the limit  $R \rightarrow \infty$.
Next,  we find the conditions under which  the integration along the half circle vanishes.
The modulus square of the integrand  of  (\ref{a501}) is given by (not including  $1/k$ that does not contribute to the integral
along the half-circle)
\barray
 \Big|   \frac{ \cosh( i k z + \pi k (L+1) )}{  \sinh( \pi k (L+1))} \Big|^2 & =  &    \frac{\rm num}{\rm den} \, ,
 \quad k = R e^{ i \theta}
 \label{a53}
 \earray
 \barray
 {\rm num} & = &
  \left[ \cosh ( R ( \cos ( \theta)  \lambda  - \sin( \theta) \tilde{x} ) ) \cos ( R ( \sin ( \theta)  \lambda  - \cos( \theta) \tilde{x} ) )   \right]^2 +
    \left[ \sinh ( R ( \cos ( \theta)  \lambda  - \sin( \theta) \tilde{x} ) ) \sin ( R ( \sin ( \theta)  \lambda  - \cos( \theta) \tilde{x} ) )   \right]^2  \, ,
\nonumber  \\
 {\rm den} & = &  \left[  \sinh( R \cosh(\theta) ) \cos( R \sin (\theta)) \right]^2  +
 \left[  \cosh( R \cosh(\theta) ) \sin( R \sin (\theta)) \right]^2  \, ,
 \nonumber
\earray
where $R$ has been replaced by $R/(\pi (L+1))$ and
\barray
\lambda &= & 1 -   \frac{ y }{ \pi (L+1)},  \qquad y = {\rm Im} \;  z,  \qquad 0 < \lambda < 1 \, ,  \label{a54} \\
\tilde{x}  &= & \frac{ x }{ \pi (L+1)},  \qquad x = {\rm Re} \;  z  \nonumber \, .
\earray
In the limit $R \rightarrow \infty$ and $\theta \neq \pi/2$,  one gets
\beq
  \frac{\rm num}{\rm den} \rightarrow {\rm exp} \left( { 2 R ( | \cos (\theta) \lambda - \sin (\theta) \tilde{x} | - |\cos \theta| }) \right)   \, ,
  \label{a55}
  \eeq
which vanishes exponentially for $\tilde{x} =0$ because $0 < \lambda < 1$, while
for $\tilde{x} \neq 0$ there is a region in $\theta$ where it diverges.
We shall then impose the condition that  $x =0$, i.e.  $z = i y$, in which case the integral (\ref{a501})
can be computed by residue calculus obtaining
\beq
I_{2,L}^\epsilon(i y) = \sum_{n=1} ^\infty \frac{ \cos( n y/(L+1))}{n}
= - \log \left( 2 \sin \frac{ y}{ 2(L+1)} \right) , \qquad 0 < y < \pi (L+1)  \, .
\label{a56}
\eeq
Similarly,  the integral (\ref{a50}) is given by  (with $z = i y$)
\beq
I_{1,L}^\epsilon(i y)
=   \log    \frac{  \sin \frac{ y}{ 2(L+1)} } { \sin \frac{ y}{ 2L}},  \qquad 0  < |y|  < \pi L   \, .
\label{a56b}
\eeq
Gathering the results obtained we have
\barray
I_{1,L}^{\epsilon}(z)  & = &  \log    \frac{  \sinh  \frac{ z}{ 2(L+1)} }{ \sinh \frac{ z}{ 2L}},
 \qquad  0 < | {\rm Im} (z)  | < \pi L   \, ,
\label{a57} \\
I_{2,L}^\epsilon(z) & = &  - \log \left( - 2  i \sinh \frac{ z}{ 2(L+1)} \right),
\qquad 0 <   {\rm Im} (z)    < \pi (L + 1)    \, ,
\nonumber  \\
I_{3,L}^\epsilon  & =  &    - \frac{1}{2} \log  \frac{L+1}{L} , \qquad  I_{4,L}^\epsilon  =   - \frac{1}{2} \log  (L+1)   \, ,
 \nonumber
\earray
 that allow us to write  eq.(\ref{a47}) as
\beq
\circled{11}  =  - \left[ \sum_{L \geq j > l \geq 1} q_j q_l  \log
 \frac{   \sinh  \frac{ z_j - z_l}{ 2(L+1)} }{ \sinh \frac{ z_j - z_l}{ 2L}} +  \sum_{j=1} ^L  q_j q_0
  \log \left( - 2  i \sinh \frac{ z_j - z_0}{ 2(L+1)} \right) - \sum_{j=1}^L    \frac{q_j^2 }{2} \log \frac{ L+1}{L}
  -     \frac{q_0^2}{2} \log (L+1)   \right]
 \label{a58}
 \eeq
 and adding (\ref{a39})
\barray
\circled{14} \equiv  \circled{1} + \circled{11} & = & - \left[ \sum_{L \geq j > l \geq 0} q_j q_l
\log  \sinh( \mu   \frac{ z_j - z_l}{ 2(L+1)} )
 - \sum_{j=1}^L    \frac{q_j^2 }{2} \log \frac{ L+1}{L}
  -     \frac{q_0^2}{2} \log (L+1)   \right] \, ,
 \label{a581}
 \earray
 where we have chosen
 \beq
 \mu = - 2 i   \, ,
 \label{a591}
 \eeq
 to simplify  the expression.
  Collecting the terms in (\ref{a451}), (\ref{a452})  and (\ref{a58}) we get
\barray
\tilde{R}_{M_1 \cup M_L} & = &
\circled{1} +   \circled{8} +  \circled{9}  + \circled{11} +   \circled{12} +  \circled{13}
\label{a59} \\
& = & \circled{8} +  \circled{9}  +    \circled{12} +   \circled{13} + \circled{14}
 \nonumber
\earray
Now,  we consider

\barray
\circled{15}  \equiv
\circled{8} +  \circled{12} &  = &  \frac{1}{2} \int_0^\infty dk \;
\left[   \hat{f}_+(k)    \hat{f}_+^*(k)  \left(  \omega_{+,1}(k)  -   \frac{  \omega_{-,1}^2(k)}{ 2 \Omega(k)} \right)   +
 \hat{f}_-(k)    \hat{f}_-^*(k)  \left(  \omega_{+,L}(k)  -   \frac{  \omega_{-,L}^2(k)}{ 2 \Omega(k)} \right)   \right.   \label{a60} \\
&  & \left.  -  ( \hat{f}_+(k)    \hat{f}_-^*(k) +  \hat{f}_-(k)    \hat{f}_+^*(k) ) \frac{  \omega_{-,L}(k) \omega_{-,L}(k) }{ 2 \Omega(k)}
\right]  \nonumber  \\
& = &   \frac{1}{2} \int_{0}^\infty     dk \;    ( \hat{f}_+(k),  \hat{f}_-(k))  \,
 \left( \begin{array}{cc}
\omega_{+,L+1}(k) &    \omega_{-,L+1}(k) \\
\omega_{-,L+1}(k) &    \omega_{+,L+1}(k) \
  \end{array}
  \right)   \left( \begin{array}{l}
  \hat{f}_+^*(k) \\
   \hat{f}_-^*(k) \\
   \end{array}
   \right)    \, ,
   \nonumber
\earray
where we used the identities
\barray
 \omega_{+,1}(k)  -   \frac{  \omega_{-,1}^2(k)}{ 2 \Omega(k)} & = &  \omega_{+,L}(k)  -   \frac{  \omega_{-,L}^2(k)}{ 2 \Omega(k)} =
 k \coth( \pi k (L+1) ) = \omega_{+, L+1}(k)  \, ,  \label{a61} \\
-   \frac{\omega_{-,L}(k) \omega_{-,L}(k) }{ 2 \Omega(k)}  & = &
 - k /\sinh( \pi k (L+1)) = \omega_{-, L+1} (k)  \, .
 \nonumber
 \earray

The other summand   in (\ref{a59}) is
\barray
\circled{16}  \equiv
\circled{9} +  \circled{13} &  = &
- \frac{i}{2} \int_\Rmath dk  \,  \frac{ q_0 e^{i k z_0}}{ \sinh( \pi k)}
\left[  \hat{f}_+(k) \left( 1 + \frac{ e^{ - \pi k} \omega_{-,1}(k) }{2 \Omega(k)}  \right)  +
\hat{f}_-(k) \frac{ e^{ - \pi k} \omega_{-,L}(k) }{2 \Omega(k)}    \right]
\label{a62} \\
&   &
+  \frac{i}{2} \int_\Rmath dk  \,  \sum_{j=1}^L  \frac{ q_j e^{i k z_j}}{ \sinh( \pi k L)}
\left[  \hat{f}_+(k) \frac{e^{  \pi k L} \omega_{-,L}(k) }{2 \Omega(k)}   +
\hat{f}_-(k) \left( 1 + \frac{ e^{  \pi k L} \omega_{-,L}(k) }{2 \Omega(k)}  \right)    \right]
\nonumber  \\
& = & - \frac{i}{2 }  \int_{\Rmath}    dk  \;   \sum_{j=0}^L   q_j    \frac{ e^{ i k z_j}}{ \sinh( \pi k (L+1))}
  ( e^{ \pi k L}  \hat{f}_+(k)-    \hat{f}_-(k) e^{- \pi k} )   \, ,
\nonumber
\earray
where we used
\barray
1 + \frac{ e^{ - \pi k} \omega_{-,1}(k) }{2 \Omega(k)} & = &  \frac{ e^{ \pi k L} \sinh( \pi k)}{ \sinh( \pi k (L+1))} ,
\qquad \frac{ e^{ - \pi k} \omega_{-,L}(k) }{2 \Omega(k)}  = -  \frac{ e^{ - \pi k } \sinh( \pi k)}{ \sinh( \pi k (L+1))} \, ,
\label{a63} \\
 \frac{e^{  \pi k L} \omega_{-,L}(k) }{2 \Omega(k)}   & = & -  \frac{ e^{ \pi k L} \sinh( \pi k L)}{ \sinh( \pi k (L+1))}, \qquad
 1 + \frac{ e^{  \pi k L} \omega_{-,L}(k) }{2 \Omega(k)}  =
 \frac{ e^{ - \pi k } \sinh( \pi k L)}{ \sinh( \pi k (L+1))}   \, .
 \nonumber
\earray
Collecting terms
\barray
\tilde{R}_{M_1 \cup M_L} & = &
\circled{14} +   \circled{15} +  \circled{16}
\label{a64}  \\
& = &   -  \sum_{L \geq j > l \geq 0} q_j q_l  \log ( \mu
 \sinh  \frac{ z_j - z_l}{ 2(L+1)}  )
 +  \sum_{j=1}^L    \frac{q_j^2 }{2} \log \frac{ L+1}{L}
  +     \frac{q_0^2}{2} \log (L+1)
  \nonumber  \\
&  & +    \frac{1}{2} \int_{0}^\infty     dk \;    ( \hat{f}_+(k),  \hat{f}_-(k))  \,
 \left( \begin{array}{cc}
\omega_{+,L+1}(k) &    \omega_{-,L+1}(k) \\
\omega_{-,L+1}(k) &    \omega_{+,L+1}(k) \
  \end{array}
  \right)   \left( \begin{array}{l}
  \hat{f}_+^*(k) \\
   \hat{f}_-^*(k) \\
   \end{array}
   \right)
   \nonumber  \\
 &   & -  \frac{i}{2 }  \int_{\Rmath}    dk  \;   \sum_{j=0}^L   q_j    \frac{ e^{ i k z_j}}{ \sinh( \pi k (L+1))}
  ( e^{ \pi k L}  \hat{f}_+(k)-    \hat{f}_-(k) e^{- \pi k} )   \, ,
\nonumber
\earray
that using   (\ref{a35}) can be written as

\barray
\tilde{R}_{M_1 \cup M_L}[f_+, f_-, \{ q_j , {z}_j \}_{j=0}^L] & =  &
R_{M_{L+1}}[f_+, f_-, \{ q_j , \tilde{z}_j \}_{j=0}^L]
\label{a65}  \\
& &
+  \sum_{j=0}^L    \frac{q_j^2 }{2} \log (L+1) -    \sum_{j=1}^L    \frac{q_j^2 }{2} \log L \, ,
\nonumber
\earray
where $\tilde{z}_j = z_j + i \pi$ takes into account that $M_1 \cup M_L = \Rmath \times [- \pi, \pi L]$
while the expression (\ref{a35}) is  associated to  the interval
 $M_{L+1} = \Rmath \times [0, \pi (L+1)]$.
The terms proportional to $q^2_j \;(j=0,1, \dots, L)$ have the following interpretation.
The functional  $A_{M_1}[  f_+, f_-, \{ z_0, q_0 \}]$
corresponds in the CFT formalism to the chiral  vertex operator $: e^{ i q_0 \varphi(z_0)}:$, which has conformal  dimension
$h = \frac{1}{2} q_0^2$ [S1].  The MPS amplitude $A_{M_L}$ should therefore  correspond to the
operator $: e^{ i \sum_{j=1}^L q_j \varphi(z_j)}:$ with  scaling dimension $\sum_{j=1}^L \frac{1}{2} q_j^2$.
This leads to the  definition of  the functional
\barray
{\cal A}_{M_{L}}[f_0, f_+, f_-, \{ q_j , z_j \}_{j=1}^L]   &  =  &   L^{ - \frac{1}{2} \sum_{j=1}^L q_j^2}  \, e^{ i f_0 \sum_{j=1}^L q_j  } \,
A_{M_L}[f_+, f_-, \{ q_j , z_j \}_{j=1}^L]  \label{a66} \\
& = &   e^{ i f_0 \sum_{j=1}^L q_j  } \,  {\rm exp} \left( - R_{M_L}[f_+, f_-, \{ q_j , z_j \}_{j=1}^L]     -   \sum_{j=1}^L \frac{ q_j^2}{2}  \log L     \right)  \, ,
\nonumber
\earray
where we have  included the zero mode $f_0$   (recall eq.(\ref{a4b})).
The sewing equation (\ref{a28}) can finally be written as
\beq
\int [d g]
{\cal A}_{M_1}[f_0, f_+, g, \{ q_0 , z_0 \}]   \;  {\cal A}_{M_L}[f_0, g, f_-, \{ q_j , z_j \}_{j=1}^L]    =
 {\cal A}_{M_1 \cup M_L}[f_0, f_+, f_-, \{ q_j , z_j \}_{j=0}^L] \, ,
\label{a67}
\eeq
that is equivalent  to the equation (5a) in the main text.

\subsection{The closing condition}

We  first  make the identification
${f}_+ = {f}_- \equiv f$ in eq. (\ref{a35}),
\barray
R_{M_L}[f, f, \{ q_j , z_j \}_{j=1}^L] & = &  -  \sum_{L \geq j > k \geq 1}   q_j q_k  \log( \mu   \sinh \frac{ z_j - z_k}{2 L} )
\label{b35} \\
&+  & \frac{1}{2} \int_{0}^\infty     dk \;    ( \hat{f}(k),  \hat{f}(k))  \,
 \left( \begin{array}{cc}
\omega_{+,L}(k) &    \omega_{-,L}(k) \\
\omega_{-,L}(k) &    \omega_{+,L}(k) \
  \end{array}
  \right)   \left( \begin{array}{l}
  \hat{f}^*(k) \\
   \hat{f}^*(k) \\
   \end{array}
   \right)
 \nonumber   \\
& -  & \frac{i}{2 }  \sum_{j=1}^L   q_j  \int_{0}^\infty     dk  \;   \left[
 \frac{ e^{ i k z_j}}{ \sinh( \pi k L)}   \hat{f}(k)  ( e^{ \pi k L} - 1)
-   \frac{ e^{-  i k z_j}}{ \sinh( \pi k L)}   \hat{f}^*(k)  ( e^{- \pi k L} - 1)  \right]
    \,
\nonumber  \\
& = &  -  \sum_{L \geq j > k \geq 1}   q_j q_k  \log( \mu   \sinh \frac{ z_j - z_k}{2 L}  )
+  \int_0^\infty dk \;  \left[
 \hat{f}(k) \hat{f}^*(k)   \tilde{\Omega}(k)   +   \hat{f}(k)   \tilde{\alpha}(k) +   \hat{f}^*(k)   \tilde{\beta}(k) \right]  \, ,
\nonumber
\earray
where
\barray
\tilde{\Omega}(k) & = & \frac{1}{2} ( \omega_{+,L}(k) + \omega_{+,L} (k)) = k  \tanh(\pi k L/2),
\label{b36} \\
\tilde{\alpha}(k) & = &  - \frac{i}{2 }
 \sum_{j=1}^L q_j  \frac{ e^{ i k z_j} ( e^{ \pi k L} -1)  }{ \sinh( \pi k L)  }   =
  - \frac{i}{2 }
 \sum_{j=1}^L q_j  \frac{ e^{ i k z_j}   e^{ \pi k L/2} }{ \cosh( \pi k L/2)  }  \, ,
 \nonumber  \\
 \tilde{\beta}(k) & = & \frac{i}{2 }
 \sum_{j=1}^L q_j  \frac{ e^{ - i k z_j} (e^{ - \pi k L} -1)  }{ \sinh( \pi k L)  }  =
-  \frac{i}{2 }
 \sum_{j=1}^L q_j  \frac{ e^{ - i k z_j} e^{ - \pi k L/2}   }{ \cosh( \pi k L/2 )  } \, ,
    \nonumber
 \earray
 and integrate over $f$  (recall eq.(\ref{a44}))
\beq
\int \prod_{k>0} d \hat{f}(k) d \hat{f}^*(k) {\rm exp} \left( - R_{M_L}[f, f, \{ q_j , z_j \}_{j=1}^L]
 \right)
  =
 {\rm exp} \left(   \sum_{L \geq j > k \geq 1}   q_j q_k  \log ( \mu   \sinh \frac{ z_j - z_k}{2 L} )
  +    \int_0^\infty  dk \frac{  \tilde{\alpha}(k) \tilde{\beta}(k)   }{  \tilde{\Omega}(k)}   \right)
 \label{b37}
  \eeq
The integral is given by
\barray
\int_0^\infty  dk \frac{  \tilde{\alpha}(k) \tilde{\beta}(k)   }{  \tilde{\Omega}(k)}
& = & - \int_0^\infty  \frac{dk}{ k}  \frac{1}{ \sinh( \pi k L)}
\left(  \sum_{L \geq j > l \geq 1} q_j q_l \cos(  k (z_j - z_l) +   \sum_{j=1}^L  \frac{ q_j^2}{2}  \right)
\label{b38}  \\
& = & - \left(   \sum_{L \geq j > l \geq 1} q_j q_l  \log \frac{   \sinh \frac{ z_j - z_l}{2 L} } {   \sinh \frac{ z_j - z_l}{ L} }
- \log 2 \sum_{j=1}^L  \frac{q_j^2}{2}   \right)
\nonumber
\earray
where we used the regularized integrals (\ref{a50}) and (\ref{a51}).
Plugging (\ref{b38}) into (\ref{b37}) gives

\barray
\int \prod_{k>0} d \hat{f}(k) d \hat{f}^*(k) {\rm exp} \left( - R_{M_L}[f, f, \{ q_j , z_j \}_{j=1}^L]
 \right)
 & =    &
 {\rm exp} \left(   \sum_{L \geq j > k \geq 1}   q_j q_k  \log ( \mu  \sinh \frac{ z_j - z_k}{ L} )
  +  \log 2 \sum_{j=1}^L  \frac{q_j^2}{2}     \right).
 \label{b39}
  \earray
This equation  together with (\ref{a66}) implies
\barray
\int [df] {\cal A}_{M_{L}}[f_0, f, f, \{ q_j , z_j \}_{j=1}^L]   & =  &   \, e^{ i f_0 \sum_{j=1}^L q_j  } \,
 {\rm exp} \left(   \sum_{L \geq j > k \geq 1}   q_j q_k  \log ( - 2 i   \sinh \frac{ z_j - z_k}{ L} )
  +  \sum_{j=1}^L  \frac{q_j^2}{2}   \log \frac{2}{L}     \right)
 \label{b40} \\
 & = & \, e^{ i f_0 \sum_{j=1}^L q_j  } \,    {\rm exp} \left(   \sum_{L \geq j > k \geq 1}   q_j q_k  \log ( - L  i   \sinh \frac{ z_j - z_k}{ L} )
  +  \frac{1}{2}  (  \sum_{j=1}^L  q_j )^2    \log \frac{2}{L}    \right)
  \nonumber   \\
  & = & \, e^{ i f_0 \sum_{j=1}^L q_j  } \,  \prod _{L \geq j > k \geq 1}    \left(   L     \sin  \frac{ y_j - y_k}{ L} \right)^{ q_j q_k}
  \left( \frac{ L}{2} \right)^{ -  \frac{1}{2}  ( \sum_{j=1}^L  q_j )^2   }     \, .
  \nonumber
 \earray
 Finally, integrating over the zero mode $f_0$ yields,
 \barray
\int df_0  [df] {\cal A}_{M_{L}}[f_0, f, f, \{ q_j , z_j \}_{j=1}^L]  &=  &
2 \pi \;  \delta( \sum_{j=1}^L q_j) \;
\prod _{L \geq j > k \geq 1}    \left(   L     \sin  \frac{ y_j - y_k}{ L} \right)^{ q_j q_k}
\label{b41} \\
  & = &  2 \pi  \;   \delta( \sum_{j=1}^L q_j) \;   \langle   \prod _{L \geq j > k \geq 1}   : e^{ i q_j \varphi(z_j) } :  : e^{ i q_k \varphi(z_k) } : \rangle_{\rm cyl}   \;
   \, .
  \nonumber
\earray
where $\langle .. \rangle_{\rm cyl}$ is the vacuum expectation value of the product of chiral  vertex operators, at positions $z_j = i y_j$
in a cylinder of length $\pi L$.  This result provides a proof of equation (5b) in the main text.

\section{PEPS functional}

\subsection{The conformal map}

\begin{figure}[h!]
\includegraphics[width=70mm]{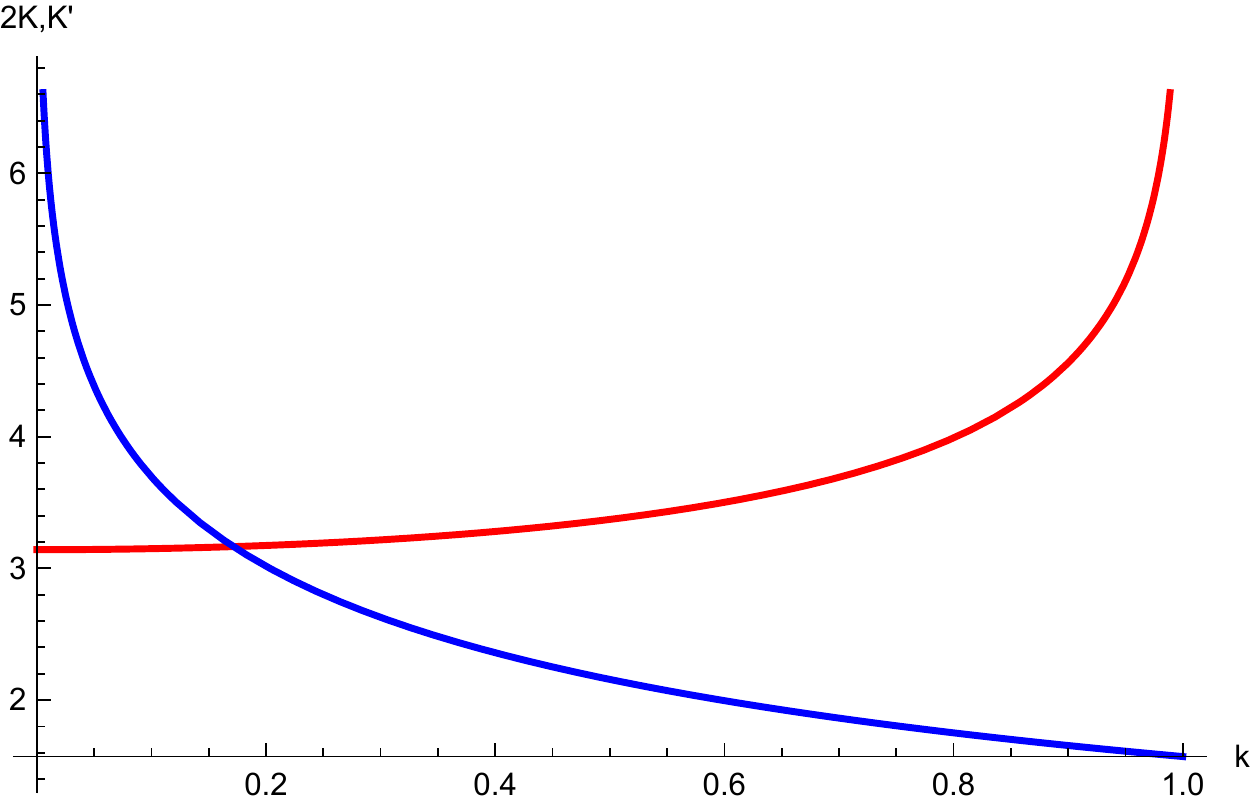}
\caption{Plot of $2 K(k)$ (red curve)  and $K'(k)$  (blue curve) as a function of the elliptic modulus $k$.
The two curves intersect at $k= 0.171573 $.}
\label{K-plot}
\end{figure}

The PEPS functional is obtained when  the region  $M$ of the path integral is a rectangle in the complex plane.
The conformal map $g: M \rightarrow \Hmath$ can be constructed using  the
Schwarz-Christoffel formula,
\beq
u=F(\sin^{-1}(z) , k) = \int_0^{z}  \frac{dt}{ \sqrt{ (1 - t^2)( 1 - k^2 t^2)}},  \quad u \in M,  \quad z \in \Hmath \, ,
\label{e4}
\eeq
where $F(\phi, k)$ is the incomplete elliptic integral  of the first kind defined as
\beq
u=F(\phi, k) = \int_0^\phi  \frac{d \theta}{\sqrt{1 - k^2  \sin^2  \theta}} = \int_0^{\sin \phi}  \frac{dt}{ \sqrt{ (1 - t^2)( 1 - k^2 t^2)}} \, ,
\label{e1}
\eeq
and $k$ is the elliptic modulus that satisfies  $0 < k^2 < 1$.
The points $z = \pm1, \pm 1/k$, on the real axis  are mapped into the vertices of the rectangle with  coordinates
\barray
z=  \pm 1 & \rightarrow &  u =   \pm  K(k) \, ,  \label{e5} \\
z=  \pm 1/k  & \rightarrow & u =   \pm  K(k) + i K'(k)  \, ,
\nonumber
\earray
where $K(k)$ is the complete elliptic integral of the first kind,
\beq
K(k) =  \int_0^{\pi/2}   \frac{d \theta}{\sqrt{1 - k^2  \sin^2  \theta}} \, ,
\label{e5b}
\eeq
 and  $K'(k) = K(k')$, with $k' = \sqrt{1-k^2}$ is the complementary modulus.
Moving along  the real axis in the $z$-plane one goes through
the points $- \frac{1}{k}, -1, 1, \frac{1}{k}$, that correspond in the
$u$-plane to the points $- K + i K', -K, K, K+ i K'$ that form the vertices
of a rectangle  of width   $2 K$  and  height   $K'$.  Figure \ref{K-plot}
shows these constants  that intersect at $k=0.171573 \dots$.
In the limit $k \rightarrow 1$ one has
\barray
\lim_{k \rightarrow 1} K(k) = \infty. \qquad  \lim_{k \rightarrow 1} K'(k) = \frac{\pi}{2}  \,  ,
\label{k1}
\earray
which shows that  the  rectangle   $M= [- K, K] \times [0, K']$ becomes  the strip $\R  \times [0, \frac{\pi}{2}]$.

 The inverse function of $F$ is  the Jacobi amplitude
 \beq
 \phi= F^{-1}(u,k)=  {\rm am}(u, k) \, ,
 \label{e2}
 \eeq
 in terms of which the Jacobi  elliptic functions are defined
\barray
\sin \phi  &  =  &   \sin (   {\rm am}(u, k)  ) = {\rm sn}(u,k)  \, ,
\label{e3} \\
\cos \phi  &  =  &   \cos (   {\rm am}(u, k)  ) = {\rm cn}(u,k)  \, ,
\nonumber  \\
\sqrt{ 1 - k^2 \sin^2 \phi} & = & \sqrt{ 1 - k^2  \sin^2 ({\rm am}(u,k))} = {\rm dn}(u,k)  \, .
\nonumber
\earray
For $k=0$ and $k=1$ the elliptic function become trigonometric and hyperbolic functions respectively
\barray
k=0 & : & sn(x) = \sin(x),  \qquad cn(x) = \cos(x), \qquad dn(x) = 1 \, ,  \label{j1}   \\
k=1 & : & sn(x) = \tanh(x),  \quad cn(x) = \frac{1}{\cosh(x)}, \quad dn(x) = \frac{1}{\cosh(x)}  \, .   \nonumber
\earray
Using  eqs. (\ref{e4})  and (\ref{e3}) the conformal map $g: M \rightarrow \Hmath$ is given by
\beq
\zeta = {\rm sn}(z, k)  \, , \qquad z \in M , \qquad \zeta \in \Hmath \, .
\label{a19}
\eeq

\subsection{The non chiral functional}

\begin{figure}[h!]
\includegraphics[width=80mm]{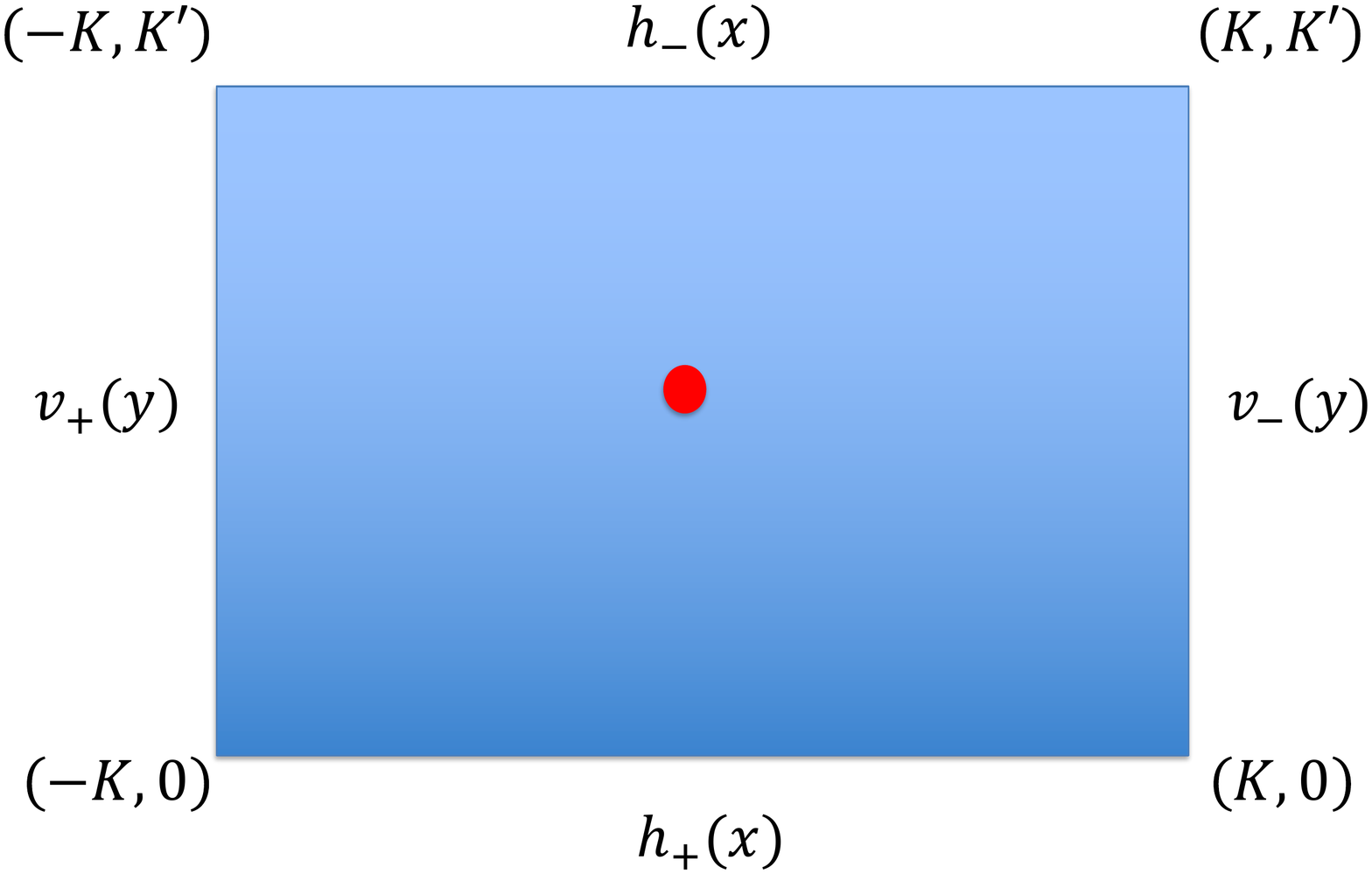}
\caption{Graphical representation of the  PEPS functional (\ref{b11first}) for $N=1$. }
\label{PEPS}
\end{figure}

The Green's function $G_M$ on $M = [- K, K] \times [0, K']$
 is obtained replacing (\ref{a19}) into (\ref{a15}),
\beq
G_{M}(z, \bar{z}; z', \bar{z}')
= \frac{1}{4 \pi}  \log  \frac{ ( sn(z) - sn(z')) ( \overline{sn(z)} -  \overline{sn(z')}  )} { ( sn(z) -  \overline{sn(z')} ( \overline{sn(z)}   - sn(z'))} \,,
\qquad  \; z, z' \in M \, ,
\label{a20}
\eeq
where $s(z) \equiv sn(z,k)$.  On the boundary of $M$  we define the functions (see fig.\ref{PEPS})
\barray
h_{+}(x)& =   &  f(x,0), \quad h_{ -}(x) = f(x, K'),  \qquad  - K \leq x \leq K \, ,  \label{b3} \\
v_{ +}(y)  & =  &  f(- K, y), \quad  v_{-}(y) = f(K,y), \qquad   0 \leq y \leq K'  \, .  \nonumber
\earray
that correspond to the functions $\alpha_{n,m}, \beta_{n,m}, \gamma_{n,m}, \delta_{n,m}$ defined in eqs. (23) in the main text.

Equation (\ref{a10}) becomes

\barray
S_{M}  &  = &  \frac{1}{8 \pi} \int_M   d^2 x  \int_M \, d^2 x' \, G_{M}( \x, \x') \rho(\x) \rho(\x')
\label{b4} \\
 & + & \frac{1}{8 \pi} \int_{-K}^K     dx    \int_{- K}^K   \, d x'  \,   ( h_{+}(x),  h_{-}(x))  \,
 \left( \begin{array}{cc}
  \partial_{y} \partial_{y'}  G_{M}|_{y_+, y_+'}    &
 -    \partial_{y} \partial_{y'}  G_{M}|_{ y_+, y_-}  \\
-   \partial_{y} \partial_{y'}  G_{M}|_{y_-, y_+} &
  \partial_{y} \partial_{y'}  G_{M}|_{y_-, y'_-} \\
  \end{array}
  \right)   \left( \begin{array}{l}
  h_{+}(x') \\
   h_{-}(x') \\
   \end{array}
   \right)
 \nonumber   \\
  & + & \frac{1}{8 \pi} \int_{0}^{K'}      dy    \int_{0}^{K'}    \, d y'  \,   ( v_+(y),  v_-(y))  \,
 \left( \begin{array}{cc}
  \partial_{x} \partial_{x'}  G_{M}|_{x_+, x_+'}    &
 -    \partial_{x} \partial_{x'}  G_{M}|_{ x_+, x_-}  \\
-   \partial_{x} \partial_{x'}  G_{M}|_{x_-, x_+} &
  \partial_{x} \partial_{x'}  G_{M}|_{x_-, x'_-} \\
  \end{array}
  \right)   \left( \begin{array}{l}
  v_+(y') \\
   v_-(y') \\
   \end{array}
   \right)
 \nonumber \\
  & + & \frac{1}{8 \pi} \int_{-K}^K     dx    \int_{0}^{K'}    \, d y'  \,   ( h_+(x),  h_-(x))  \,
 \left( \begin{array}{cc}
  \partial_{y} \partial_{x'}  G_{M}|_{y_+, x_+}    &
 -    \partial_{y} \partial_{x'}  G_{M}|_{ y_+, x_-}  \\
-   \partial_{y} \partial_{x'}  G_{M}|_{y_-, x_+} &
  \partial_{y} \partial_{x'}  G_{M}|_{y_-, x_-} \\
  \end{array}
  \right)   \left( \begin{array}{l}
  v_+(y') \\
   v_-(y') \\
   \end{array}
   \right)
 \nonumber
 \\
  & + & \frac{1}{8 \pi} \int_{0}^{K'}      dy    \int_{- K}^K   \, d x'  \,   ( v_+(y),  v_-(y))  \,
 \left( \begin{array}{cc}
  \partial_{x} \partial_{y'}  G_{M}|_{x_+, y_+}    &
 -    \partial_{x} \partial_{y'}  G_{M}|_{ x_+, y_-}  \\
-   \partial_{x} \partial_{y'}  G_{M}|_{x_-, y_+} &
  \partial_{x} \partial_{y'}  G_{M}|_{x_-, y_-} \\
  \end{array}
  \right)   \left( \begin{array}{l}
  h_+(x') \\
   h_-(x') \\
   \end{array}
   \right)
 \nonumber
 \\
 & + & \frac{1}{4 \pi} \int_{-K}^{K}     dx    \int_M \, d^2  x'  \,  \rho(\x')   ( h_+(x),  h_-(x))
   \left(
\begin{array}{c}
  \partial_{y}  G_{M}(x, y; x' , y')|_{y_+} \\
-   \partial_{y}  G_{M}(x, y; x' , y')|_{y_-}\\
\end{array}
  \right)  \,
\nonumber
 \\
 & + & \frac{1}{4 \pi} \int_{0}^{K'}     dx    \int_M \, d^2  x'  \,  \rho(\x')   ( v_+(y),  v_-(y))
   \left(
\begin{array}{c}
  \partial_{x}  G_{M}(x, y; x' , y')|_{x_+} \\
-   \partial_{x}  G_{M}(x, y; x' , y')|_{x_-}\\
\end{array}
  \right)  \, ,
\nonumber
\earray
where
\beq
x_+ = - K + \varepsilon, \quad x_- = K - \varepsilon, \quad  y_+ = \varepsilon, \quad  y_- = K' - \varepsilon \, \quad \varepsilon > 0,
\label{b5}
\eeq
and the corresponding primed versions as in eq.(\ref{a223}) needed  to regularize  the expressions.
 Replacing $z= x + i y$ and $z' = x' + i y'$ into (\ref{a20}) we get
\barray
G_{M}(x, y; x', y')
&= &  \frac{1}{4 \pi}  \log  \frac{ ( sn(x+ i y) - sn(x' + i y')) ( sn(x - i y)  - sn(x' - i y')  )}
{  ( sn(x+ i y) - sn(x' - i y')) ( sn(x - i y)  - sn(x' + i y')  )} \,,  \label{b6}
%\\
\earray
which is defined in the rectangle
\barray
   \;  - K \leq x, x' \leq K, \quad 0 \leq y, y' \leq K'  \, .
\label{b7}
\earray
Since $k$ is a real parameter, we used that $\overline{sn(x+i y)}  = sn(x-i y)$.
Taking the derivatives  respect to $x$ and $y$ and using
\beq
\frac{ d \, sn(z)}{dz} = cn(z) dn(z)
\label{b8}
\eeq
one obtains
\barray
\partial_y G_{M}(x, y; x', y')
&= &  \frac{i}{4 \pi} \left( \frac{ cn(x+ i y) dn(x+ i y)}{ sn(x+ i y) - sn(x' + i y')} -   \frac{ cn(x- i y) dn(x- i y)}{ sn(x- i y) - sn(x' - i y')}  \right.
\label{b9}  \\
&  & \left. -   \frac{ cn(x+ i y) dn(x+ i y)}{ sn(x+ i y) - sn(x' -  i y')} +    \frac{ cn(x- i y) dn(x- i y)}{ sn(x- i y) - sn(x' + i y')}  \right)   \, ,
\nonumber \\
\partial_x G_{M}(x, y; x', y')
&= &  \frac{1}{4 \pi} \left( \frac{ cn(x+ i y) dn(x+ i y)}{ sn(x+ i y) - sn(x' + i y')} +  \frac{ cn(x- i y) dn(x- i y)}{ sn(x- i y) - sn(x' - i y')}  \right.
\nonumber   \\
&  & \left. -   \frac{ cn(x+ i y) dn(x+ i y)}{ sn(x+ i y) - sn(x' -  i y')} -   \frac{ cn(x- i y) dn(x- i y)}{ sn(x- i y) - sn(x' + i y')}  \right)   \, ,
\nonumber
\earray
and
\barray
\partial_y  \partial_{y'}  G_{M}(x, y; x', y')
&= &  - \frac{1}{4 \pi}  \left(
\frac{ cn(x+ i y) dn(x+ i y) cn(x'+ i y') dn(x'+ i y')}{ (sn(x+ i y) - sn(x' + i y'))^2} +
\frac{ cn(x- i y) dn(x- i y) cn(x'- i y') dn(x'- i y')}{( sn(x- i y) - sn(x' - i y'))^2}  \right.
\nonumber   \\
&  & \left. +
\frac{ cn(x+ i y) dn(x+ i y) cn(x'-  i y') dn(x'-  i y') }{ (sn(x+ i y) - sn(x' -  i y'))^2} +
\frac{ cn(x- i y) dn(x- i y) cn(x'+ i y') dn(x'+ i y') }{ (sn(x- i y) - sn(x' + i y'))^2}  \right)   \, ,
\nonumber \\
\partial_x  \partial_{x'}  G_{M}(x, y; x', y')
&= &  \frac{1}{4 \pi}  \left(
\frac{ cn(x+ i y) dn(x+ i y) cn(x'+ i y') dn(x'+ i y')}{ (sn(x+ i y) - sn(x' + i y'))^2} +
\frac{ cn(x- i y) dn(x- i y) cn(x'- i y') dn(x'- i y')}{( sn(x- i y) - sn(x' - i y'))^2}  \right.
\nonumber   \\
&  & \left. -
\frac{ cn(x+ i y) dn(x+ i y) cn(x'-  i y') dn(x'-  i y') }{ (sn(x+ i y) - sn(x' -  i y'))^2} -
\frac{ cn(x- i y) dn(x- i y) cn(x'+ i y') dn(x'+ i y') }{ (sn(x- i y) - sn(x' + i y'))^2}  \right)   \, ,
\nonumber   \\
\partial_y  \partial_{x'}  G_{M}(x, y; x', y')
&= &  \frac{i}{4 \pi}  \left(
\frac{ cn(x+ i y) dn(x+ i y) cn(x'+ i y') dn(x'+ i y')}{ (sn(x+ i y) - sn(x' + i y'))^2}
- \frac{ cn(x- i y) dn(x- i y) cn(x'- i y') dn(x'- i y')}{( sn(x- i y) - sn(x' - i y'))^2}  \right.
\nonumber   \\
&  & \left.
- \frac{ cn(x+ i y) dn(x+ i y) cn(x'-  i y') dn(x'-  i y') }{ (sn(x+ i y) - sn(x' -  i y'))^2} +
\frac{ cn(x- i y) dn(x- i y) cn(x'+ i y') dn(x'+ i y') }{ (sn(x- i y) - sn(x' + i y'))^2}  \right)   \, ,
\nonumber \\
\partial_x  \partial_{y'}  G_{M}(x, y; x', y')
&= &  \frac{i}{4 \pi}  \left(
\frac{ cn(x+ i y) dn(x+ i y) cn(x'+ i y') dn(x'+ i y')}{ (sn(x+ i y) - sn(x' + i y'))^2}
-\frac{ cn(x- i y) dn(x- i y) cn(x'- i y') dn(x'- i y')}{( sn(x- i y) - sn(x' - i y'))^2}  \right.
\nonumber   \\
&  & \left. +
\frac{ cn(x+ i y) dn(x+ i y) cn(x'-  i y') dn(x'-  i y') }{ (sn(x+ i y) - sn(x' -  i y'))^2} -
\frac{ cn(x- i y) dn(x- i y) cn(x'+ i y') dn(x'+ i y') }{ (sn(x- i y) - sn(x' + i y'))^2}  \right)   \, .
\nonumber
\earray

Using (\ref{b5}) together with (see [S2])
\barray
sn(-z) & = &  - sn(z), \quad cn(-z) = cn(z), \quad dn(-z) = dn(z) \, ,  \label{b10}  \\
sn(z \pm K) & = &  \pm \frac{cn(z)}{dn(z)}, \quad cn(z \pm K) = \mp \frac{ k' sn(z)}{dn(z)}, \quad dn(z \pm K) = \frac{ k'}{dn(z) } \, ,  \nonumber  \\
sn(z  \pm  i K') & = &   \frac{1}{k  \, sn(z)}, \quad cn(z \pm i K') = \mp i \frac{   dn(z)}{k \,  sn(z)}, \quad dn(z \pm i  K') = \mp i  \frac{cn(z)}{sn(z) }  \, ,
\nonumber  \\
sn(i y |k) & = & i \frac{ sn(y, k')}{ cn(y |k')} , \quad cn(i y |k) =  \frac{1}{ cn(y |k')} , \quad dn(i y |k) =  \frac{ dn(y, k')}{ cn(y |k')}
\, ,
\nonumber
\earray
we derive
\barray
S_{M}  &  = &  \frac{1}{8 \pi} \int_M   d^2 x  \int_M \, d^2 x' \, G_{M}( \x, \x') \rho(\x) \rho(\x')
\label{b11first} \\
 & -  &  \frac{1}{64 \pi^2} \int_{-K}^K     dx    \int_{- K}^K   \, d x'  \,   ( h_{+}(x),  h_{-}(x))  \,
 \left( \begin{array}{cc}
U_{h_+, h_+}(x, x')    &
    U_{h_+, h_-}(x, x')  \\
 U_{h_-, h_+}(x, x')  &
   U_{h_-, h_-}(x, x')  \\
  \end{array}
  \right)   \left( \begin{array}{l}
  h_{+}(x') \\
   h_{-}(x') \\
   \end{array}
   \right)
 \nonumber   \\
  & - & \frac{1}{64 \pi^2} \int_{0}^{K'}      dy    \int_{0}^{K'}    \, d y'  \,   ( v_+(y),  v_-(y))  \,
 \left( \begin{array}{cc}
U_{v_+, v_+}(y, y')     &
{U}_{v_+, v_-}(y, y')    \\
{U}_{v_-, v_+}(y, y')   &
{U}_{v_-, v_-}(y, y')  \\
  \end{array}
  \right)   \left( \begin{array}{l}
  v_+(y') \\
   v_-(y') \\
   \end{array}
   \right)
 \nonumber \\
  & - & \frac{1}{64 \pi^2} \int_{-K}^K     dx    \int_{0}^{K'}    \, d y'  \,   ( h_+(x),  h_-(x))  \,
 \left( \begin{array}{cc}
U_{h_+, v+}(x,y')     &   U_{h_+, v_-}(x,y')  \\
U_{h_-, v+}(x,y')&  U_{h_-, v_-}(x,y')  \\
  \end{array}
  \right)   \left( \begin{array}{l}
  v_+(y') \\
   v_-(y') \\
   \end{array}
   \right)
 \nonumber \\
 & + & \frac{i}{16 \pi^2}  \int_{-K}^{K}     dx    \int_M \, d^2  x'  \,  \rho(\x')   ( h_+(x),  h_-(x))
   \left(
\begin{array}{r}
 V_{h_+} (x, {\bf x'}) -   \overline{ V_{h_+} (x, {\bf x'}) }   \\
-  V_{h_-} (x, {\bf x'}) +    \overline{ V_{h_-} (x, {\bf x'}) }  \\
\end{array}
  \right)  \,
\nonumber
 \\
 & + & \frac{i}{16 \pi^2} \int_{0}^{K'}     dy     \int_M \, d^2  x'  \,  \rho(\x')   ( v_+(y),  v_-(y))
   \left(
\begin{array}{r}
-  V_{v_+} (y, {\bf x'}) +    \overline{ V_{v_+} (y, {\bf x'}) }   \\
 V_{v_-} (y, {\bf x'}) -   \overline{ V_{h_-} (y, {\bf x'}) }  \\
\end{array}
  \right)  \, ,
\nonumber
\earray
where
\barray
U_{h_+,h_+}(x, x') & = & U_{h_-,h_-}(x, x')  =  2   cn(x) dn(x) cn(x') dn(x')
 \left( \frac{ 1}{ (sn(x+ i \varepsilon) - sn(x' + i \varepsilon'))^2}   \right.
\label{b11b}    \\
&  &  \left.
+ \frac{1}{( sn(x- i \varepsilon) - sn(x' - i \varepsilon'))^2}
+
\frac{ 1 }{ (sn(x+ i \varepsilon) - sn(x' -  i \varepsilon'))^2} +
\frac{ 1 }{ (sn(x- i \varepsilon) - sn(x' + i \varepsilon'))^2}  \right)   \, ,
\nonumber \\
%%%%%%%%%%%%%%%%%%%%%
& & \nonumber \\
U_{h_+, h_-}(x, x') & = & U_{h_-, h_+}(x, x') =  8 k \frac{ cn(x) dn(x) cn(x') dn(x')}{ ( 1- k^2 sn(x) sn(x'))^2} \, ,  \nonumber \\
%%%%%%%%%%%%%%%%%%%%%
%%%%%%%%%%%%%%%%%%%%%
{U}_{v_+, v_+ }(y, y') & = & {U}_{v_-, v_- }(y, y') =  2 k'^4\, \widetilde{cn}(y) \widetilde{sn}(y) \widetilde{cn}(y')  \widetilde{sn}(y')  \left(
\frac{ 1}{ (\widetilde{dn}(y+ i \varepsilon) -  \widetilde{dn}(y' + i \varepsilon'))^2}
 \right.
\nonumber   \\
& & \nonumber \\
&  & \left. +
\frac{1}{( \widetilde{dn}(y- i \varepsilon) -  \widetilde{dn}(y' - i \varepsilon'))^2}  +
\frac{ 1 }{ ( \widetilde{dn}(y+ i \varepsilon) -  \widetilde{dn}(y' -  i \varepsilon'))^2} +
\frac{ 1 }{ ( \widetilde{dn}(y- i \varepsilon) -  \widetilde{dn}(y' + i \varepsilon'))^2}  \right)   \, ,
\nonumber \\
%%%%%%%%%%%%%%%%%%%%%
\vspace{0.2cm}
& & \nonumber \\
\vspace{0.2cm}
{U}_{v_+, v_-}(y, y') & = & {U}_{v_-, v_+}(y, y')  =  8 k'^4  \,  \frac{ \widetilde{cn} (y)  \widetilde{sn}(y)   \widetilde{cn}(y')  \widetilde{ sn}(y')}{ ( \widetilde{dn}(y) + \widetilde{dn}(y')   )^2}  \, , \nonumber  \\
%%%%%%%%%%%%%%%%%%%%%
\vspace{0.2cm}
& & \nonumber \\
\vspace{0.2cm}
U_{h_+, v_+}(x, y') & = & 16 \, k'^2 \,  \frac{ cn(x) dn(x) \widetilde{sn}(y') \widetilde{dn}(y')}{ ( 1 + sn(x) \widetilde{dn}(y') )^2} ,
\qquad U_{h_-, v_-}(x, y')  =  16 \,  k k'^2 \,  \frac{ cn(x) dn(x) \widetilde{cn}(y') \widetilde{sn}(y')}{ ( k \,  sn(x) -  \widetilde{dn}(y') )^2} \, ,
\nonumber \\
\vspace{0.2cm}
& & \nonumber \\
\vspace{0.2cm}
U_{h_+, v_-}(x, y') & = & 16 \,  k'^2 \,  \frac{ cn(x) dn(x) \widetilde{cn}(y') \widetilde{sn}(y')}{ (1- sn(x)  \widetilde{dn}(y') )^2} ,
\qquad  U_{h_-, v_+}(x, y')  =     16  \,  k  k'^2 \,  \frac{ cn(x) dn(x) \widetilde{sn}(y') \widetilde{dn}(y')}{( k \,  sn(x) +  \widetilde{dn}(y') )^2}  \, ,
 \nonumber  \\
\vspace{0.2cm}
V_{h_+}(x, {\bf x'}) & = &     \frac{ 2 \,  cn(x) dn(x) }{  sn(x) -  sn(x' +  i y') }, \qquad  \qquad  \qquad
V_{h_-}(x, {\bf x'})  =      \frac{ 2 \,  cn(x) dn(x) }{ sn (x) (  1  - k \,   sn  (x) \,    sn(x' +  i y') )} \, ,
  \nonumber  \\
\vspace{0.2cm}
V_{v_+}(x, {\bf x'}) & = &      \frac{ 2 \, k'^2  \,  \widetilde{sn}(y)  \widetilde{cn}(y) }{  \widetilde{dn}(y) ( 1 + \widetilde{dn}(y)   sn(x' +  i y') )}  ,
\qquad
V_{v_-}(x, {\bf x'})  =      \frac{ 2 \, k'^2  \,  \widetilde{sn}(y)  \widetilde{cn}(y) }{  \widetilde{dn}(y) ( 1 -  \widetilde{dn}(y)   sn(x' +  i y') )}  \, ,
 \nonumber
\earray
and
\beq
\widetilde{sn}(y) = sn(y |k'),  \quad \widetilde{cn}(y) = cn(y |k'),  \quad \widetilde{dn}(y) = dn(y |k'),  \quad k'^2 = 1 - k^2 \, .
\label{b12}
\eeq
The regularize versions of $U_{h_\pm,h_\pm}$ and $U_{v_\pm,v_\pm}$ is given in
eqs.(27) of the main text.

The  chiral version of (\ref{b11first}) that we propose is

\barray
R_{M}  &  = &  - \sum_{N \geq j > k \geq 1} q_j q_k  \ln  (sn(z_j ) - sn(z_k))
\label{b11second} \\
 & -  &  \frac{1}{64 \pi^2} \int_{-K}^K     dx    \int_{- K}^K   \, d x'  \,   ( h_{+}(x),  h_{-}(x))  \,
 \left( \begin{array}{cc}
U_{h_+, h_+}(x, x')    &
    U_{h_+, h_-}(x, x')  \\
 U_{h_-, h_+}(x, x')  &
   U_{h_-, h_-}(x, x')  \\
  \end{array}
  \right)   \left( \begin{array}{l}
  h_{+}(x') \\
   h_{-}(x') \\
   \end{array}
   \right)
 \nonumber   \\
  & - & \frac{1}{64 \pi^2} \int_{0}^{K'}      dy    \int_{0}^{K'}    \, d y'  \,   ( v_+(y),  v_-(y))  \,
 \left( \begin{array}{cc}
U_{v_+, v_+}(y, y')     &
{U}_{v_+, v_-}(y, y')    \\
{U}_{v_-, v_+}(y, y')   &
{U}_{v_-, v_-}(y, y')  \\
  \end{array}
  \right)   \left( \begin{array}{l}
  v_+(y') \\
   v_-(y') \\
   \end{array}
   \right)
 \nonumber \\
  & - & \frac{1}{64 \pi^2} \int_{-K}^K     dx    \int_{0}^{K'}    \, d y'  \,   ( h_+(x),  h_-(x))  \,
 \left( \begin{array}{cc}
U_{h_+, v+}(x,y')     &   U_{h_+, v_-}(x,y')  \\
U_{h_-, v+}(x,y')&  U_{h_-, v_-}(x,y')  \\
  \end{array}
  \right)   \left( \begin{array}{l}
  v_+(y') \\
   v_-(y') \\
   \end{array}
   \right)
 \nonumber \\
 & - & \frac{1}{ 4 \pi }  \sum_{j=1}^N    \int_{-K}^{K}     dx  \; q_j   ( h_+(x),  h_-(x))
   \left(
\begin{array}{r}
 V_{h_+} (x, z_j )    \\
-  V_{h_-} (x, z_j )   \\
\end{array}
  \right)  \,
\nonumber
 \\
 & -  & \frac{1}{4 \pi}  \sum_{j=1}^N  \int_{0}^{K'}    dy \;   q_j    ( v_+(y),  v_-(y))
   \left(
\begin{array}{r}
-  V_{v_+} (y, z_j)   \\
 V_{v_-} (y, z_j)   \\
\end{array}
  \right)  \, .
\nonumber
\earray
where we have used the charge density (\ref{a12}).

Equations (\ref{k1}) shows  that in the limit $k \rightarrow 1$, the rectangle
$M= [- K, K] \times [0, K']$ becomes the strip $\R  \times [0, \frac{\pi}{2}]$.
Therefore,  the PEPS functional  (\ref{b11second}) must be closely related to  the MPS functional  (\ref{a26})
with  $a=0$ and $\Delta = 1/2$.  The reason for this fact  is the following.
In the limit  $k \rightarrow 1$,  the conformal map (\ref{a19}) becomes

\beq
g_1(z) = \tanh(z) = \frac{ e^{2 z} -1}{ e^{2 z} +1}
\label{b14}
\eeq
where we used (\ref{j1}).  On the other hand, the conformal map (\ref{a201}),
with $a=0, \Delta = 1/2$, is $g(z) = e^{ 2 z}$. Notice that $g_1(z)$ is
a M{\"o}bius transformation of $g(z)$, so we expect the wave functions
constructed with both functionals to be the same. This issue will be considered elsewhere in more detail.

\vspace{10mm}

* On leave from Department of Physics and Astronomy, Aarhus University, DK-8000 Aarhus C, Denmark

\noindent [S1] P. Di Francesco, P. Mathieu, D. S\'en\'echal, {\em Conformal Field Theory}, Springer, New York, 1997.

\noindent [S2] M.  Abramowitz and I. Stegun, {\em Handbook of mathematical functions}, Dover Publications, Inc. New York 1972.

\end{document}